\documentclass{article}  

\newcommand{\apjl}{Astrophys. J. Lett.}   
\newcommand{\apjs}{Astrophys. J. Suppl. Ser.}   
\newcommand{\ao}{Appl. Opt.}   
\newcommand{\aap}{Astron. Astrophys.}   

\usepackage{stfloats} 
\usepackage{graphicx}
\usepackage[a4paper, total={6in, 8in}]{geometry}

\newcommand*\arcsec{\ensuremath{^{\prime\prime}}}

\usepackage{amsmath}
 
\usepackage{txfonts} 
\usepackage{hyperref}
\hypersetup{
    colorlinks = True,
}
\usepackage{listings}
\usepackage{xcolor}
\usepackage{subfiles}
\usepackage[normalem]{ulem}

\usepackage{pifont}
\newcommand{\cmark}{\ding{51}}%
\newcommand{\xmark}{\ding{55}}%

\begin{document} 

\title{Instrument-To-Instrument translation: Instrumental advances drive restoration of solar observation series via deep learning}

\author{
R. Jarolim$^1$* \and 
A. M. Veronig$^{1, 2}$ \and 
W. P\"otzi$^2$ \and 
T. Podladchikova$^3$
}

\date{%
    $^1$University of Graz, Institute of Physics, Universitätsplatz 5, 8010 Graz, Austria\\%
    $^2$University of Graz, Kanzelhöhe Observatory for Solar and Environmental Research, Kanzelhöhe 19, 9521 Treffen am Ossiacher See, Austria\\%
    $^3$Skolkovo Institute of Science and Technology, Bolshoy Boulevard 30, bld. 1, Moscow 121205, Russia\\%
}

    \maketitle
    \begin{abstract}  
    The constant improvement of astronomical instrumentation provides the foundation for scientific discoveries. In general, these improvements have only implications forward in time, while previous observations do not benefit from this trend. 
    Here we provide a general deep learning method that translates between image domains of different instruments (Instrument-To-Instrument translation; ITI). We demonstrate that the available data sets can directly profit from the most recent instrumental improvements, by applying our method to five different applications of ground- and space-based solar observations. We obtain 1) solar full-disk observations with unprecedented spatial resolution, 2) a homogeneous data series of 24 years of space-based observations of the solar EUV corona and magnetic field, 3) real-time mitigation of atmospheric degradations in ground-based observations, 4) a uniform series of ground-based H$\alpha$ observations starting from 1973, 5) magnetic field estimates from the solar far-side based on EUV imagery. The direct comparison to simultaneous high-quality observations shows that our method produces images that are perceptually similar and match the reference image distribution.
    \end{abstract}

%



\section{Introduction}
\label{section:introduction}

With the rapid improvement of space-based and ground-based solar observations, unprecedented details of the solar surface and atmospheric layers have been obtained. As compared to the 11-year solar activity cycle the development of new instruments progresses over smaller time scales. This imposes additional challenges for the study of long-term variations and combined usage of different instruments. In this study, we address the question on how the information obtained from the most recent observations can be utilized to enhance observations of lower quality. This especially aims at homogenizing long time-series, mitigating atmospheric influences and to overcome instrumental limitations (e.g., resolution limitation, seeing).
 
The automated homogenization of data sets provides an integral component for long-term studies (e.g., solar cycle studies, historic sunspots, studies of rare events) and for studies that combine data from multiple instruments. Especially when dealing with large amounts of data, the automatic adjustment can be faster and more consistent than treating the data sets of the individual instruments separately \cite{hamada2020new, hamada2021chseries}. Data driven methods rely on the diversity and amount of data, and the inclusion of additional data sets can significantly increase the performance \cite{goodfellow2016deep}. Enhancing old observation series (e.g., recorded on film) to the standard and quality of the primary modern data set can provide an easily accessible data source. Methods developed for specific instruments often depend on certain observables (e.g., magnetograms, filtergrams of specific wavelengths) that are only partially covered by other data sets. An approximation based on proxies can already provide a suitable basis for automated methods or gives additional information for solar monitoring (e.g., in the frame of space weather predictions). In the regular observation schedule the mitigation of atmospheric effects and quality enhancement is a frequently addressed problem \cite{ramos2018real, woger2008speckle, rimelle2011solaradaptive}.

A principle problem of every enhancement method is the absence of a reference high-quality image. The inversion of the image degradation (e.g., lower spatial resolution, instrumental characteristics) is therefore inferred from artificial degradations \cite{schawinski2017galaxygan, rahman2020super}, from simulation data \cite{baso2018enhancing, jia2019cycle_multifractal, baso2019solar} or by estimating the degrading effects \cite{herbel2018psf, ramos2021wavefront}. We argue that a designed degradation can only represent a limited set of observations and can not account for the full diversity of real quality decreasing effects that occur in solar observations. Especially when dealing with atmospheric effects (i.e., clouds, seeing) and instrumental characteristics, the quality degradation is complex to model \cite{jarolim2020image, santos2021aiacalibration}. Even with the precise knowledge about the degrading function, every image enhancement problem is ill-posed \cite{borman1998super, yang2019deep}. We argue that we can reduce the number of possible high-quality solutions for a given low-quality image significantly by considering the distribution of real high-quality images and requiring that enhanced images correspond to the same image domain.


In this study, we propose an approach that uses real observations from state-of-the-art instruments in observational solar physics, to model the image quality distribution of high-quality images.  We overcome the limitation of a high-quality reference image with the use of unpaired image-to-image translation \cite{zhu2017unpaired}. We provide a general method that translates from a given low-quality domain to a target high-quality domain (Instrument-To-Instrument translation; ITI). With this approach, we infer information from real observations to enhance physically relevant features which are otherwise beyond the diffraction limit of the telescope (e.g., super resolution), inter-calibrate data sets, mitigate atmospheric degradation effects and estimate observables that are not covered by the instrument.

\begin{figure}%
\includegraphics[width=\linewidth]{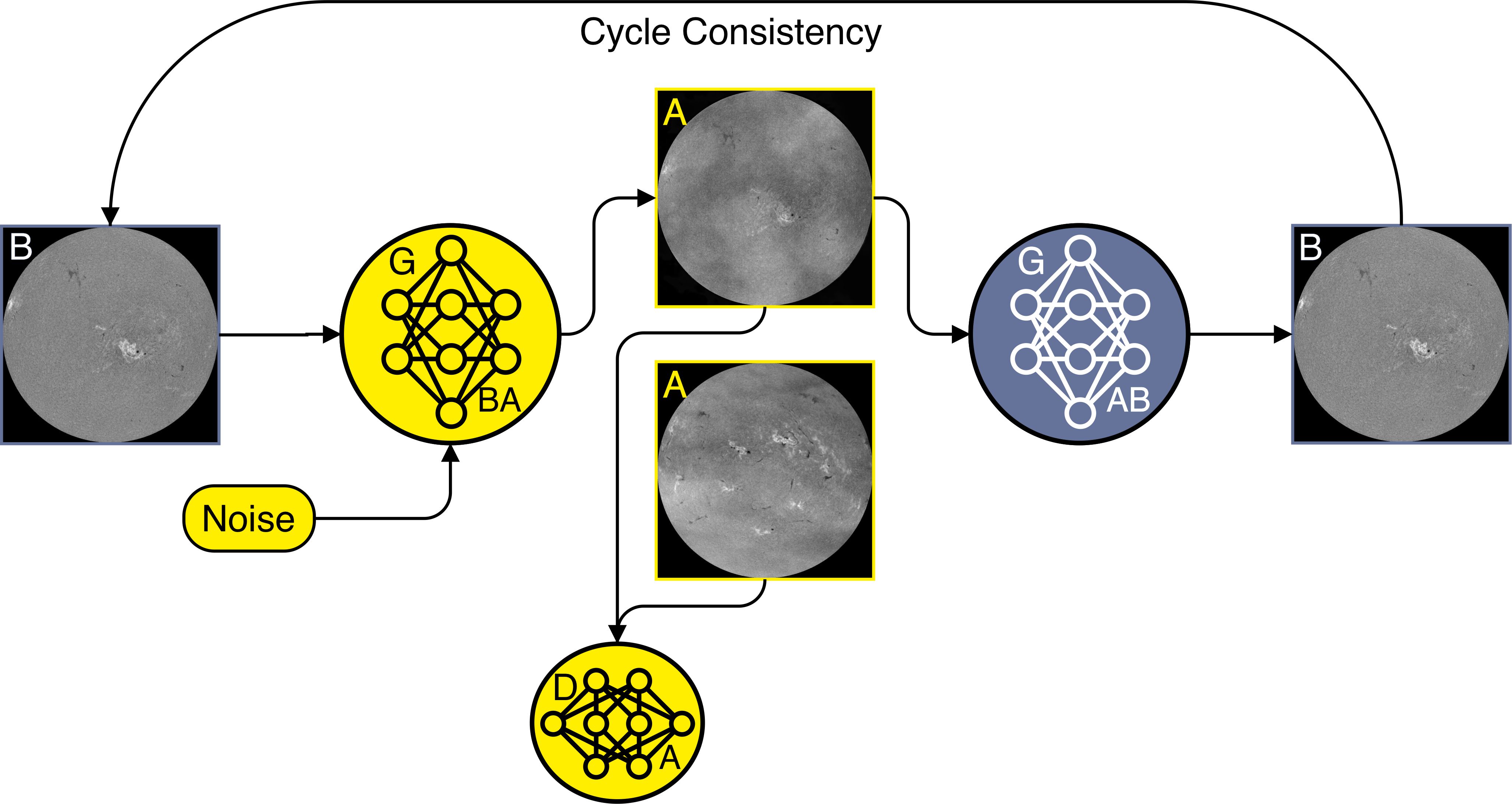}
\caption{Model training cycle for the synthesis of low-quality images. Images are transformed from the high-quality domain (B) to the low-quality domain (A) by generator BA (yellow). The synthetic images are translated by generator AB (blue) back to domain B. The mapping into domain A is enforced by discriminator A, which is trained to distinguish between real images of domain A (bottom) and generated images (top). Both generators are trained jointly to fulfill the cycle consistency between original and reconstructed image, as well as for the generation of synthetic images that correspond to domain A. The generation of multiple low-quality versions from a single high-quality image is  accomplished with the additional noise term that is added to generator BA.}
\label{figure:iti_BAB}
\end{figure}

Our primary model architecture consists of two neural networks, where the first generates synthetic low-quality images from a given high-quality image (\textit{generator BA}). The second network is trained to invert the image degradation to reconstruct the original high-quality observation (\textit{generator AB}). We enforce the generation of low-quality images with the use of competitive training between generator BA and a discriminator network. We include an additional noise factor for generator BA to model a variety of degrading effects, independent of the image content \cite{zhu2017multimodel, huang2018multimodal}. With the synthesis of more realistic and diverse low-quality observations, the generator AB is capable to provide a similar reconstruction performance for real low-quality observations (Fig. \ref{figure:iti_BAB}). The artificial degradation leads inevitably to an information loss that needs to be compensated by the generator AB to reconstruct the original image. Analogously to the training cycle in Fig. \ref{figure:iti_BAB} we employ a cycle translating low-quality observations to high-quality observations (A-B-A, Sect. \ref{section:methods}). This enforces that images by generator AB correspond to the domain of high-quality images, restricting the possible enhanced solutions and gaining information from the high-quality image distribution.

In contrast to recently proposed paired image translation tasks in solar physics \cite{park2019mag_to_euv, shin2020caII_to_mag, jeong2020ai_mag_pfss, lim2021euv_to_euv, son2021euv_to_he}, our method requires no alignment of the data sets. This enables the translation between instruments, that have no joined observing periods (e.g., historic data, different observing times, differences in cadence), or have differences in their spatial alignment (e.g., different field-of-view, observations from different vantage points in the heliosphere).

\section{Results}
\label{section:results}

In this study, we address the validity of the enhanced images with the use of five data set pairs and the assessment of the sparse set of simultaneous observations. We use ITI to obtain 1) solar full-disk observations with unprecedented spatial resolution, 2) a homogeneous data series of 24 years of space-based observations of the solar EUV corona and magnetic field, 3) real-time mitigation of atmospheric degradations in ground-based observations, 4) a uniform series of ground-based H$\alpha$ observations starting from 1973, that comprises solar observations recorded on photographic film and CCD, 5) magnetic field estimates from the solar far-side based on multi-band EUV imagery.

A description of the used instruments, data set preparation, preprocessing and training parameters can be found in the supplementary materials (App. \ref{section:data_set}, \ref{section:preprocessing}). Real high-quality observations that are used as reference are separated by a temporal split from the training data set (see App. \ref{section:preprocessing}), to exclude a potential memorization of observations.

For all applications, we distinguish between high- and low-quality data sets. Hereby we consider data set pairs, where the high-quality data set contains observations with a better spatial resolution or less quality degradations (e.g., noise, atmospheric effects), as compared to the low-quality data set. As an example, the HMI instrument onboard SDO provides high resolution observations of the Sun, but is defined as low-quality data set as compared to high-resolution observations of Hinode/SOT.

\subsection{Image super resolution with different field-of-view -\newline SDO/HMI-to-Hinode/SOT}
\label{section:hmi_hinode}

The Solar Optical Telescope onboard the Hinode satellite (Hinode/SOT; \cite{tsuneta2008sot}) provides a large data set of partial-Sun continuum images, similar to the full-disk continuum images of the Helioseismic and Mangetic Imager (HMI; \cite{schou2012hmi}) onboard the Solar Dynamics Observatory (SDO: \cite{pesnell2012sdo}). The observations by Hinode/SOT cover various regions of the Sun which makes them suitable as high-quality target for the enhancement of HMI continuum observations.

Using unpaired image translation we do not require a spatial or temporal overlap between the data sets, moreover the model training is performed with small patches of the full images (Sect.~\ref{section:methods}). This enables the use of instruments that can observe only a fraction of the Sun for the enhancement of full-disk observations. 

Here, we resize Hinode observations to 0.15 arcsec pixels and use ITI to super resolve HMI observations by a factor of 4. Hinode/SOT provides a spatial sampling of up to 0.0541 arcsec pixels, for our application we found that a resolution increase by a factor of 4 is already at the limit of where we can properly assess enhanced details.

\begin{figure}%
\includegraphics[width=\linewidth]{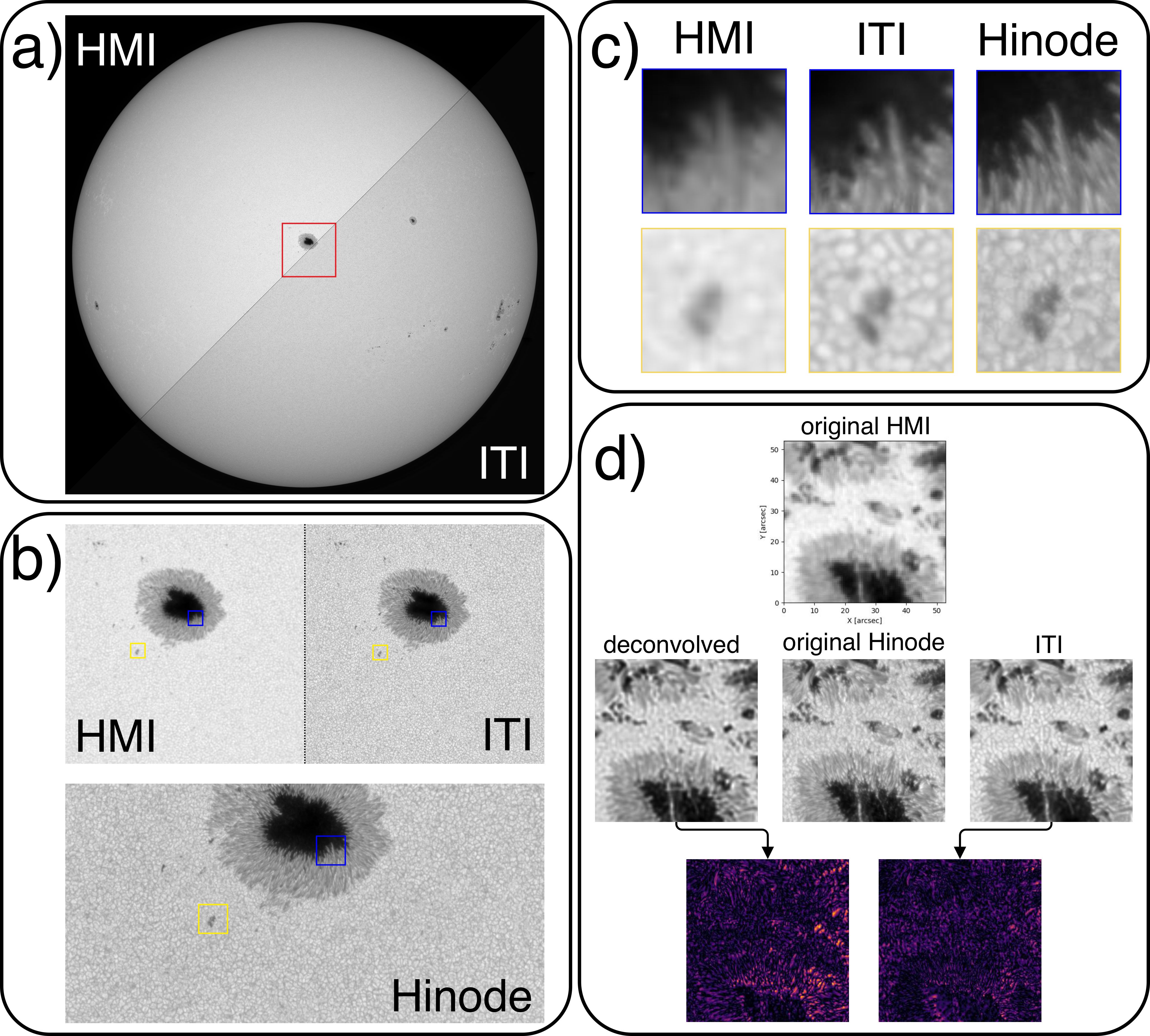}
\centering
\caption{Comparison of simultaneous observations of SDO/HMI and Hinode/SOT continuum from 2013-11-18 17:46:20 and 2013-11-18 17:46:28, respectively. a) Original HMI (top left) and ITI enhanced (bottom right) full-disk observation. b) Comparison of a $222\arcsec\times222\arcsec$ region and corresponding Hinode observation with full field-of-view and resolution. c) Matched features of the original HMI, ITI enhanced and real Hinode observation with $13\arcsec\times13\arcsec$ spatial extend. The penumbral features of ITI match the real observation. The small scale separation of the individual fibrils is beyond the resolution limit of HMI, but are largely reconstructed in the ITI observation. The coarse shape of the granulation pattern and the solar pore match the Hinode observation, while smaller intergranular lanes show deviations. d) Example of the comparison between ITI and the deconvolution approach (2014-12-19 22:33). We show the original HMI image, the enhanced images and the difference maps to the original Hinode image. The ITI image shows a better agreement with the high-resolution reference, as can be seen from the reduced errors in the penumbra and sharper boundaries of the pores.}
\label{figure:hmi_hinode_full_disk}
\end{figure}

In Fig. \ref{figure:hmi_hinode_full_disk} we show an example of the full-disk HMI and ITI enhanced HMI observations (panel a). For a direct comparison, we manually extract two subframes from the HMI full-disk images and the Hinode/SOT image (panel b). The blue box shows the umbra and penumbra of the observed sunspot. The direct comparison shows that the enhanced version is close to the actual observation and correctly separates fibrils that are only observed as blurred structure in the original observation. The yellow box shows a pore and the surrounding solar granulation pattern. Here, we can observe an enhancement of the shape that is close to the Hinode/SOT observation. The granulation pattern is similar to the Hinode/SOT observations for larger granule, but shows differences in terms of shapes and inter granular lanes. Comparing these results to the original HMI observation, the deviations result mostly from diffraction limited regions that are not related to more extended solar features (e.g., extended fibrils, coherent granulation pattern). The correct generation of fibrils in the penumbra, that are beyond the resolution limit of HMI observations, show that the neural network correctly learned to infer information from the high-quality image distribution. The clear structures in the granulation pattern can be interpreted as a result of the perceptual optimization (training cycle ABA).

Both instruments show a substantial overlap that allows for a quantitative assessment of our method. Here, we compare our method to a state-of-the-art Richardson-Lucy deconvolution \cite{richardson192deconv}, using the point-spread-function of the HMI telescope from \cite{yeo2014hmi_psf}, and calibrating the pixel counts to the Hinode scale (Sect. \ref{section:align_hmi}). We select all observations that contain activity features in the months November and December, which were excluded from the training, and acquire the observations at the closest time instance from HMI ($<$10s difference; total of 59 observations). For the pixel-wise comparison we register the images based on the phase-correlation of the Fourier-transformed images \cite{reddy1996fftregister}. For this evaluation, we focus on the similarity of features, where we normalize the image patches such that scaling offsets are ignored. In Fig.~\ref{figure:hmi_hinode_full_disk}d we show an example for paired image patches. Both the deconvolved image and the ITI enhanced image show a general increase in perceptual quality. At smaller scales the ITI image shows a better agreement with the Hinode reference image. This can be best seen from the difference images, that show a smaller deviation of the penumbra and sharper boundaries of pores than for the deconvolved image. In contrast, the quiet-Sun region shows a slightly increased error for ITI, which can be interpreted as a mismatch of the enhanced granules. The evaluation of the full test set shows that the ITI enhanced images provide throughout a higher similarity to the Hinode observations. The structural similarity index measure (SSIM) increases from 0.55 for the original images to 0.64 for the ITI enhanced images, while the images deconvolved with the Richardson-Lucy method result in 0.59. The largest improvement is achieved for the perceptual quality, as estimated by the FID, where the distance between the image distributions decreases by 40\% for the ITI images. For this metric the deconvolved images provide no improvement (-6\%).

In Supplementary Table \ref{table:hmi_comparison} we summarize the full evaluation. Additional samples of the aligned data set are given in App. \ref{section:app_hmi_hinode} and movies of full-cadence ITI enhanced observations are provided online (Movie 1 and 2).

\subsection{Re-calibration of multi-instrument data -\newline SOHO/EIT- and STEREO/EUVI-to-SDO/AIA}
\label{section:stereo_soho_sdo}

With the translation between image domains our method can account for both image enhancement and adjustment of the instrumental characteristics simultaneously (cf. \cite{ignatov2018wespe}). We use ITI to enhance EUV images from the Solar and Heliospheric Observatory (SOHO; \cite{domingo1995soho}) and Solar Terrestrial Relations Observatory (STEREO; \cite{kaiser2008stereo}) to SDO quality and calibrate the images into a unified series dating back to 1996.

With a spatial sampling of 0.6 arcsec pixels, we consider the filtergrams of the Atmospheric Imaging Assembly (AIA; \cite{lemen2012aia}) and the magnetograms of HMI onboard SDO as high-quality reference. We use the EUV filtergrams from the Exteme Ultraviolet Imager (EUVI; \cite{wulser2004euvi}) onboard STEREO as low-quality data set. For the translation of SOHO observations we use filtergrams of the Extreme-ultraviolet Imaging Telescope (EIT; \cite{delaboudiniere1995eit}) in combination with the LOS magnetograms of the Michelson Doppler Imager (MDI; \cite{scherrer1995mdi}). The SDO images provide a 2.7 times higher pixel resolution than the STEREO/EUVI images. We reduce the resolution of STEREO observations to 1024$\times$1024 pixels, before using ITI to increase the resolution by a factor 4, to the full 4096$\times$4096 pixels resolution of SDO (STEREO/EUVI-to-SDO/AIA). For observations from \mbox{SOHO/EIT+MDI} we use half of the SDO resolution as reference (SOHO/EIT+MDI-to-SDO/AIA+HMI). All observations are normalized to a fixed scale of the solar radius, to avoid variations due to the ecliptic orbits (App. \ref{section:preprocessing}). We consider all matching filters (131~$\AA$, 195/193~$\AA$, 284/211~$\AA$, 304~$\AA$, LOS magnetogram) and translate the combined set of channels with ITI, to benefit from the inter-channel relation. 

\begin{figure}%
\centering
\includegraphics[width=0.5\linewidth]{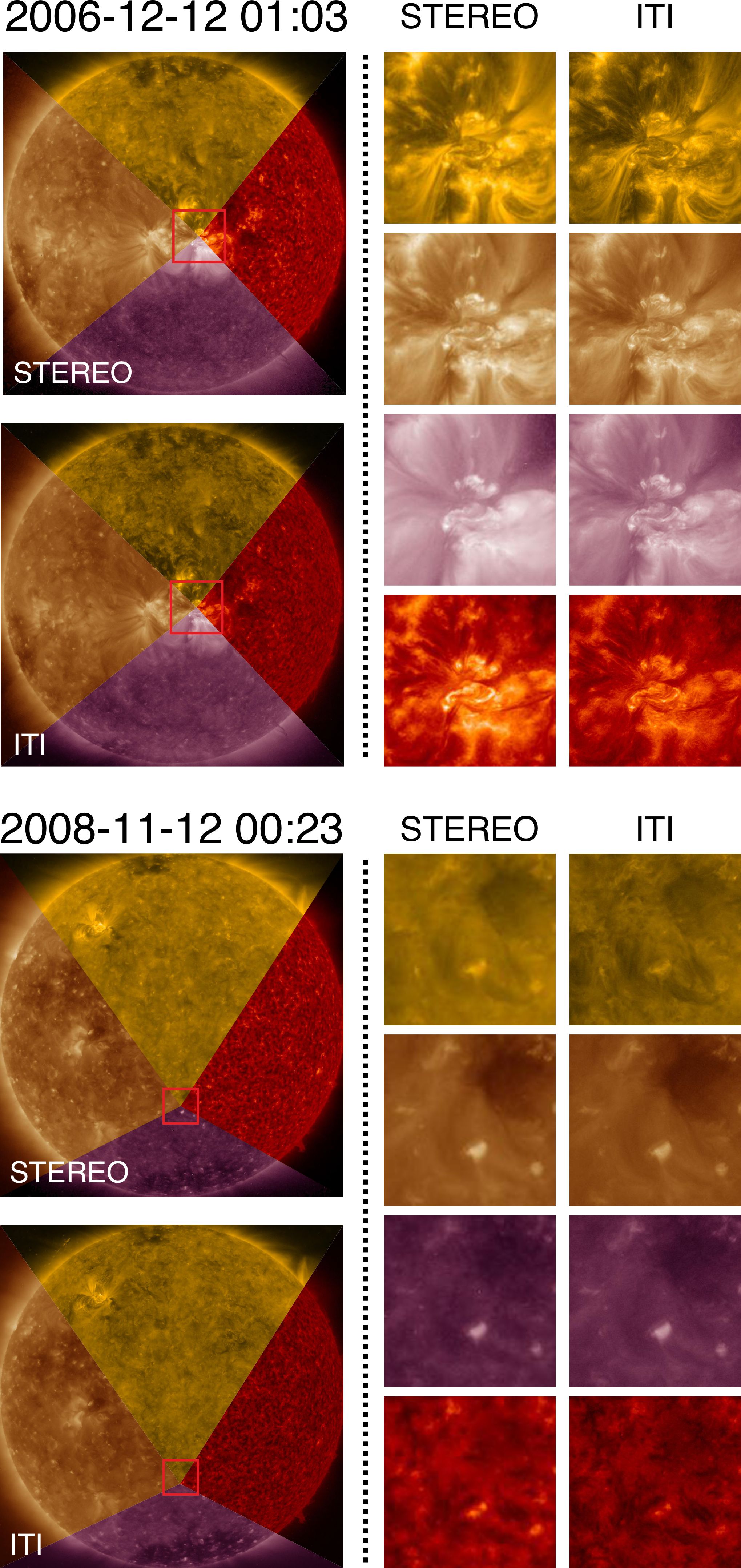}
\caption{Comparison of STEREO/EUVI and ITI enhanced EUV filtergrams. From top to bottom we show the 131~$\AA$, 195/193~$\AA$, 284/211~$\AA$ and 304~$\AA$ filtergrams. The sample at the top and bottom are taken during solar maximum and minimum, respectively. We note an overall improvement in sharpness by ITI, that can be best seen from the coronal loops of the active region (131~$\AA$ top). From the quiet-Sun region (bottom) the enhancement of faint details can be seen, with the largest perceptual quality increase for the 304~$\AA$ channel.}
\label{figure:stereo_comparison}
\end{figure}

The STEREO mission provides stereoscopic observations of the Sun, with the satellites being positioned at different vantage points. This excludes a direct comparison of the ITI enhanced observations with the SDO/AIA images. Here, we assess the usability of ITI to merge observations from multiple vantage points. In Fig. \ref{figure:stereo_comparison} we show two examples of enhanced STEREO observations. The calibration effect results from a feature dependent adjustment, where the observational characteristics of the high-quality instrument are reproduced.
The ITI enhanced observations show an improvement of sharpness, that can be best seen from the coronal loops that are observed at the resolution limit of STEREO, and appear better resolved in the ITI image. The comparison of the 284/211 $\AA$ and 304 $\AA$ channels shows that additional structures are obtained by ITI that can not be seen in the original observation. We associate this enhancement with an inference of information from the other wavelength channels, by projecting the joined set of channels to the high-quality domain.
 
\begin{figure}%
\centering
\includegraphics[width=\linewidth]{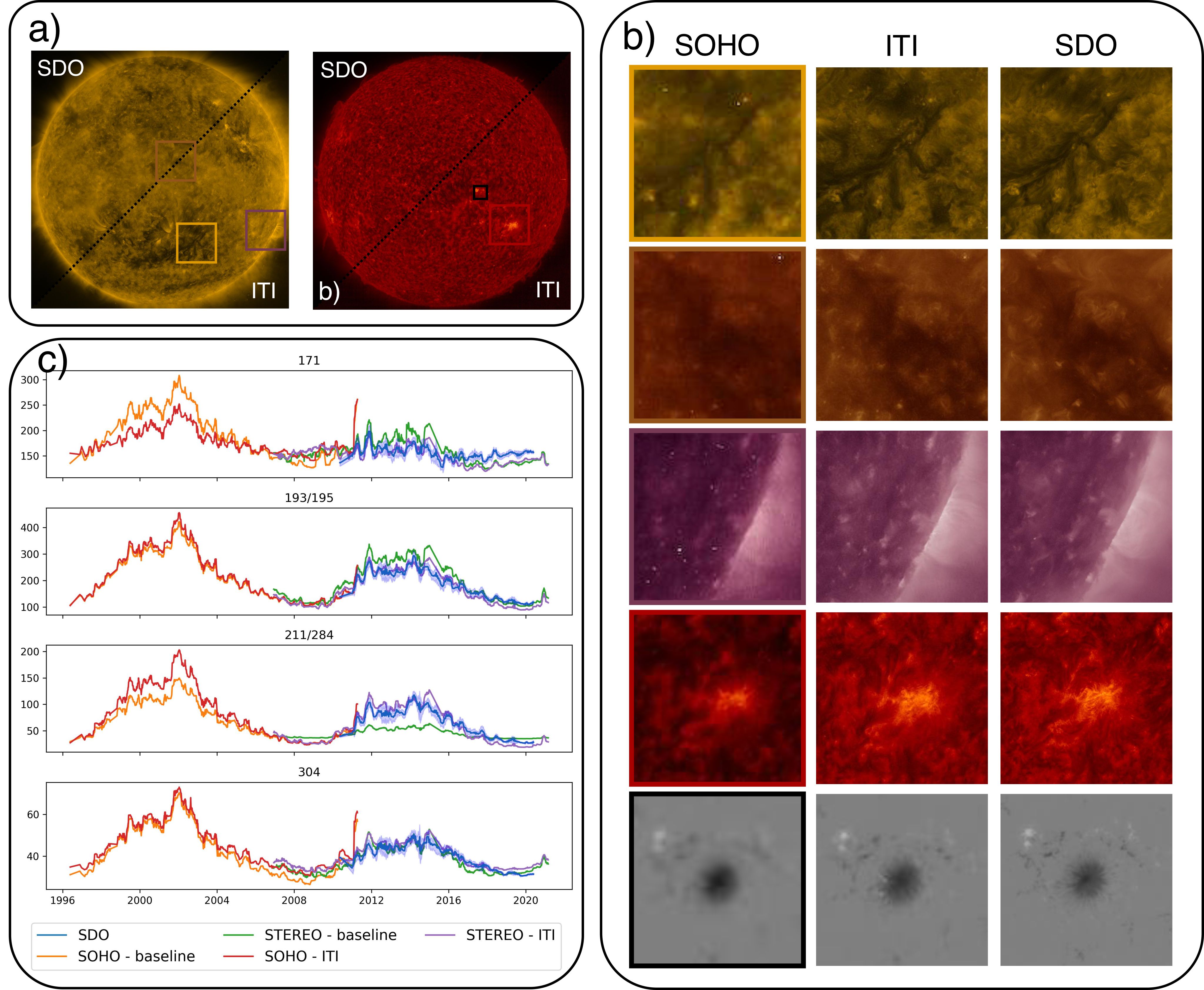}
\caption{ITI translation for the homogenization of SOHO with SDO observations and calibration of EUV data series. a) Two examples of SOHO-to-SDO translation from 2010-05-13 07:00 (left) and 2010-08-02 01:00 (right). The boxes indicate the cutouts in panel b) where we compare ITI images with aligned SDO/AIA filtergrams. We compare $300\arcsec\times300\arcsec$ and $100\arcsec\times100\arcsec$ cutouts of EUV filtergrams and LOS magnetograms, respectively. We directly compare observations taken at the same time from SOHO (left), ITI (center) and SDO (right). 131~$\AA$: The observed feature is difficult to identify in the SOHO observation, while the ITI enhanced version resolves a clear filament structure that is consistent with the SDO observation. 195/193~$\AA$: details in the quiet-Sun region are completely blurred in the SOHO image. The obtained features by ITI are consistent on a global scale with SDO, but more deviations occur at the smallest resolution scales. 284/211~$\AA$: ITI recovers faint off-limb loops that are not resolved by SOHO. The pixel-noise at the bottom left is mitigated, but results in spurious features. 304~$\AA$: The active region shows a valid reconstruction from the strongly pixelated observation. Magnetogram: Small magnetic elements are better resolved and appear deconvolved in the ITI image. The shape of the sunspot is well reconstructed, however the full quality of the SDO/HMI magnetograms can not be reached by ITI. c) Comparison of the SDO reference and calibrated SOHO/EIT and STEREO/EUVI EUV light-curves. The mean intensities for each channel are plotted against time in the individual panels. The data is smoothed by a running average filter, where the blue shaded area corresponds to 1 standard deviation of the SDO intensities. The solar cycle trend can be seen for each light-curve \cite{sidc}. ITI adjusts the individual observations to a similar scale (DN/pix/s), which outperforms the baseline approach.}
\label{figure:soho_combined}
\end{figure}

The SOHO mission provides observations that partially overlap with observations from SDO. In Fig. \ref{figure:soho_combined}b, we provide a side-by-side comparison between the original SOHO, enhanced ITI and reference SDO observations. The full-disk images are a composite of the ITI enhanced and SDO observations, which show the same calibration effect as the STEREO maps (Fig. \ref{figure:soho_combined}a). The sub-frames show the four EUV filtergrams and the LOS magnetogram of different regions of interest. We show examples of a filament, a quiet-Sun region, the solar limb and an active region. For all samples we note a strong resolution increase and high similarity to the real reference observations. The filament in the top row, is difficult to observe in the original SOHO observation, but can be clearly identified in the ITI image. For the frame at the solar-limb in the 284/211$\AA$ channel, we also note an inference from the multi-channel information that reconstructs the faint off-limb loops. From the quiet-Sun region (second row) and the full-disk images, we can see that our method is consistent across the full solar disk. The observation of the 304 $\AA$ channel (active region; fourth row) shows the strongest improvement as compared to the original observation, but also shows that smaller features could not be fully reconstructed. We note that pixel errors can lead to wrong translations and can be accounted for prior to the translation (e.g., noise in SOHO/EIT 284~$\AA$). The magnetic elements in the LOS magnetogram of ITI show an improvement in sharpness and the reconstructed shape of the sunspot matches the HMI reference, but the full HMI quality cannot be reached. The magnetic flux elements at the lowest resolution level were not resolved by ITI. This shows the limit of the method, but also suggests that no artificial features are added by our model.
 
For the usage of long-term data sets the consistent calibration is of major importance \cite{boerner2014photometric, santos2021aiacalibration}. Data sets that comprise multi-instrument data require an adjustment into a uniform series \cite{hamada2020new, chatzistergos2019analysis}. We evaluate the model performance for long-term consistency over more than two solar cycles by computing the mean intensity per channel and instrument and comparing the resulting light-curves, where we use a running average with a window size of one month to mitigate effects of the different positions of the SDO and STEREO satellites (Fig. \ref{figure:soho_combined}c). We compare our method to a baseline approach where we calibrate the EUV observations based on the quiet-Sun regions (Sect. \ref{section:euv_baseline}). As can be seen from Fig. \ref{figure:soho_combined}c, our model correctly scales the SOHO/EIT and STEREO/EUVI observations to the SDO intensity scale, with a good overlap that suggest also valid calibrations for pre-SDO times. At the end of the nominal operation of SOHO a sudden increase in the 304~$\AA$ channel can be noticed, this change in calibration restricts further assessment. Prior to this intensity jump, we note a good agreement between ITI and SDO/AIA 304~$\AA$. In Supplementary Table \ref{table:euv_calibration} we quantify the differences to the SDO reference. For the STEREO observations our method shows throughout a higher performance, especially for the coronal EUV emission lines (171 $\AA$, 195 $\AA$, 284 $\AA$). The SOHO comparison covers only simultaneous observations, which were recorded during solar minimum. Consequently, the baseline shows little deviations, and similar results are achieved by ITI.

\begin{figure}%
\centering
\includegraphics[width=\linewidth]{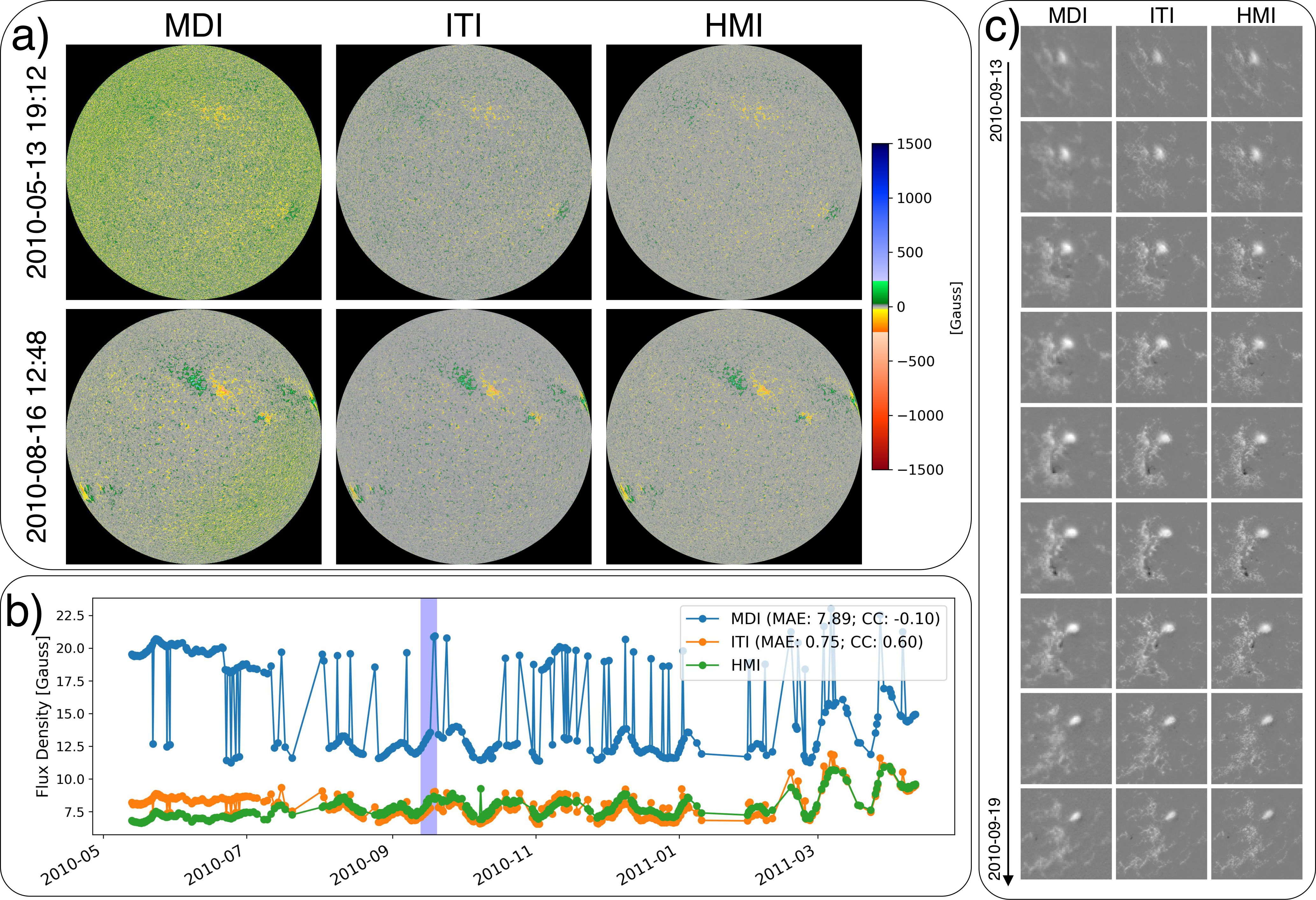}
\caption{Comparison of simultaneously observed SOHO/MDI, ITI enhanced, and SDO/HMI LOS magnetograms. a) Visual comparison of the global calibration differences. The MDI magnetograms show global inhomogeneities, that prevent a simple calibration of the series. ITI largely mitigates these variations, leading to more consistent observations, that are in agreement with the HMI reference. b) Unsigned magnetic flux density of the three magnetogram series for the period of parallel MDI and HMI observations. The MDI series shows substantial fluctuations, that are mitigated by ITI. The absolute error (MAE) and cross-correlation (CC) evaluations indicate a strong improvement in similarity between the ITI adjusted series and the HMI reference. c) Direct comparison of the active region NOAA 11106 over seven consecutive days. The LOS magnetograms cover $150\arcsec\times150\arcsec$ and are scaled by the maximum flux density in the map ($\pm 1800$ Gauss). ITI images show a similar appearance and sharpness to the SDO/HMI observations, while being also consistent with the low-resolution SOHO/MDI observations. The blue shaded area in panel b indicates the time frame of the images shown.
}
\label{figure:soho_mag}
\end{figure}

In the time between March 2010 and April 2011, magnetograms of both SOHO/MDI and SDO/HMI can be directly compared. Here we compare the calibration between MDI and HMI. As can be seen from Fig. \ref{figure:soho_mag}b the MDI instrument has an offset to the measured SDO/HMI flux. In addition, fluctuations prevent a simple calibration of the MDI magnetograms. For the considered time period, the evaluation of the MDI data shows an offset by $\sim8$ Gauss (MAE) and no correlation between the data series (correlation-coefficient of -0.1). The application of ITI leads to an intercalibration of the LOS magnetograms that accounts both for the offset (MAE of 0.8 Gauss) and mitigates the differences in scaling (correlation-coefficient of 0.6), leading to a more consistent data series. For a direct comparison we track the active region NOAA 11106 from 2010-09-13 to 2010-09-19. In Fig.~\ref{figure:soho_mag} we compare SOHO/MDI magnetograms and enhanced ITI magnetograms to the SDO/HMI magnetic field measurements. The ITI magnetograms show a clear increase in sharpness and better identification of magnetic elements as compared to the SOHO/MDI data. Differences to the reference magnetogram occur at the smallest scales, where the SOHO/MDI resolution limit is reached, while the overall flux distribution is in good agreement.

\subsection{Mitigation of atmospheric effects -\newline KSO H$\alpha$ low-to-high quality}
\label{section:kso}

The mitigation of atmospheric effects in ground-based solar observations imposes two major challenges for enhancement methods. (1) Since observations are obtained from a single instrument, there exists no high-quality reference for a given degraded observation. (2) The diversity of degrading effects is large, which commonly leads to reconstruction methods that can only account for a subset of degrading effects.

We use ITI to overcome these challenges by translating between the image domains of clear observations (high-quality) and observations that suffer from atmospheric degradations (low-quality) of the same instrument. We use ground-based full-disk H$\alpha$ filtergrams from Kanzelh\"ohe Observatory for Solar and Environmental Research (KSO; \cite{potzi2021kso}) that we automatically separate into  high- and low-quality observations (cf. \cite{jarolim2020image}). For all observations we apply a center-to-limb variation (CLV) correction and resize the observations to 1024x1024 pixels (see App. \ref{section:preprocessing}). In the resulting low-quality data set we include regular low-quality observations (e.g., clouds, seeing), while we exclude strong degradations (e.g., instrumental errors, strong cloud coverage).

\begin{figure}%
\includegraphics[width=\linewidth]{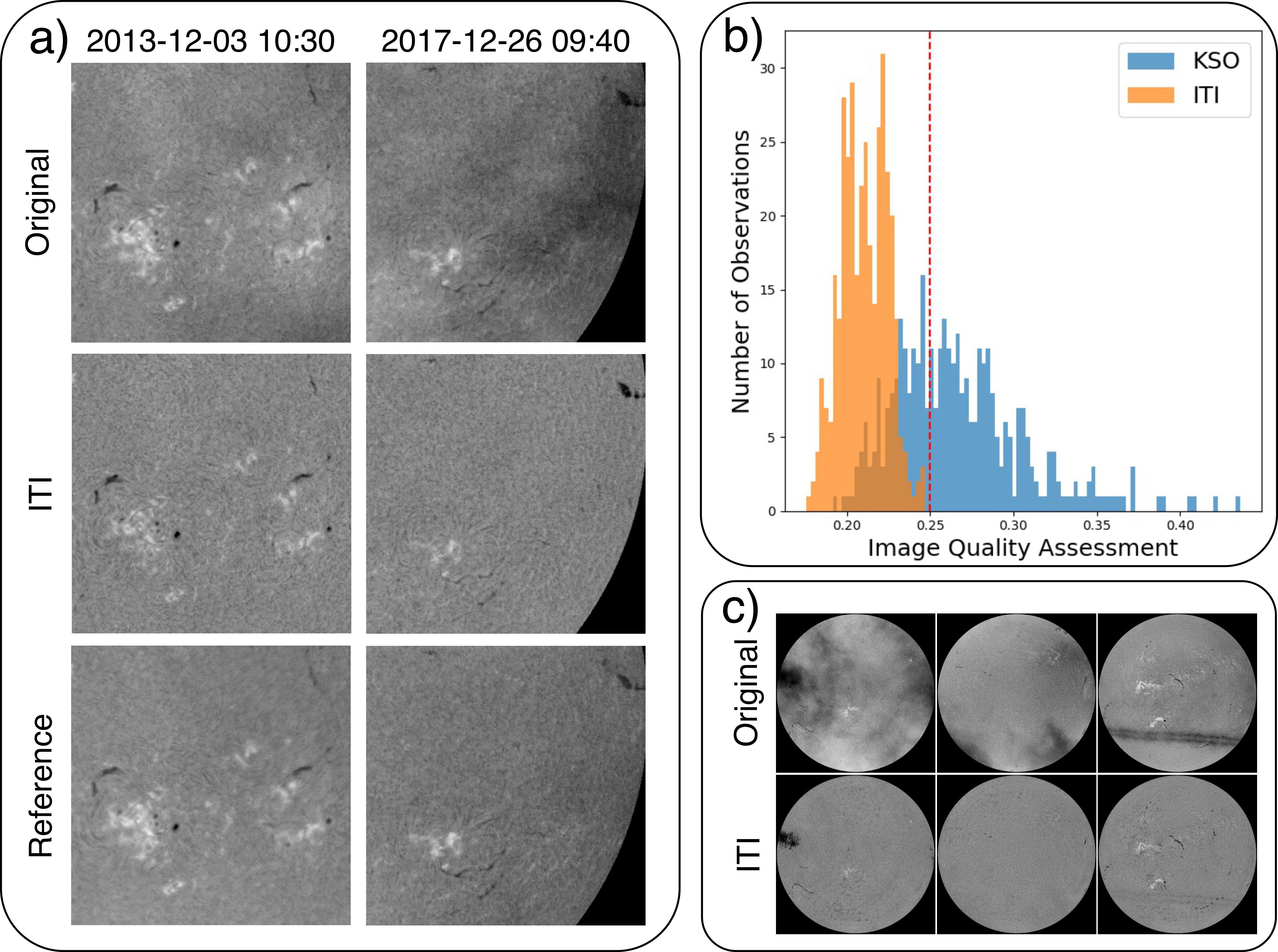}
\caption{ITI translation for the mitigation of atmospheric degradations. a) Comparison of the ITI mitigation of atmospheric effects during varying observing conditions. From top to bottom we show low-quality KSO observations, the ITI reconstructed observations and reference high-quality observations that were taken minutes after the low-quality observation. We show two samples ($800\arcsec \times 800\arcsec$), where clouds are present in the low-quality KSO observation. The ITI reconstruction leads to clearer and unobstructed observations, where small chromospheric features remain unchanged and appear sharper. b) Estimated image quality distribution of the original low-quality KSO observations (blue) and the ITI enhanced observations (orange). The red dashed line indicates the 0.25 quality threshold. c) Three full-disk images with the lowest image quality after the ITI enhancement ($\sim 0.24$).}
\label{figure:kso_combined}
\end{figure}

In Fig. \ref{figure:kso_combined}a we show two samples where a low-quality and high-quality observation were obtained within a 15 minutes time frame. We use these samples for a direct qualitative comparison of the ITI enhanced images (middle row Fig. \ref{figure:kso_combined}a), which show a removal of clouds and are in agreement with the real high-quality observation.

We use the manually selected low-quality observation from Jarolim \textit{et al.} \cite{jarolim2020image} to provide a quantitative evaluation of the quality improvement. We adapt the image quality metric by Jarolim \textit{et al.} \cite{jarolim2020image} for CLV corrected observations and only consider high-quality observations for model training, such that the quality metric estimates the deviation from the preprocessed synoptic KSO observations. In Fig. \ref{figure:kso_combined}b we show the quality distribution of low-quality KSO observations and the ITI enhanced observations. The distributions show that ITI achieves a general quality increase, where the mean quality improves from 0.27 to 0.21, for the considered data set. In Fig. \ref{figure:kso_combined}c we show three samples where the quality value intersect with the low-quality distribution after the ITI enhancement. The samples show that dense clouds and the contrail can be strongly reduced but not fully removed, which leads to the reduced image quality.

\subsection{Restoration of long time-series -\newline KSO H$\alpha$ Film-to-CCD}
\label{section:film}

\begin{figure}%
\centering
\includegraphics[width=\linewidth]{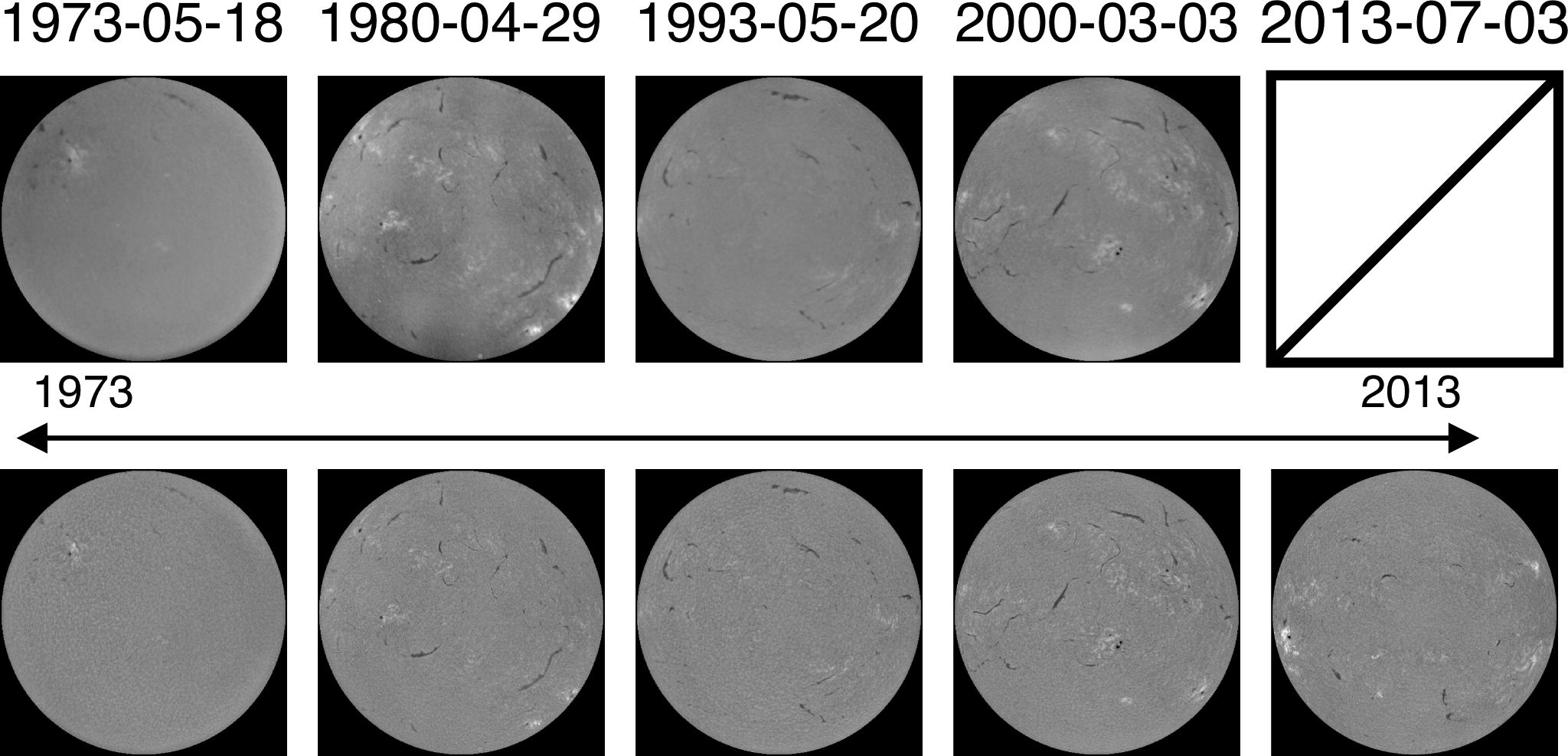}
\caption{Series of the observations on photographic film (top) and the restored ITI series (bottom). The observation on 2013-07-03 corresponds to a real high-quality KSO observation. The ITI images form a homogeneous series with the KSO archive past 2010.}
\label{figure:film_series}
\end{figure}

Instrumental upgrades inevitably lead to significant difference in data series and require adjustment in order to use combined data series \cite{chatzistergos2019analysis}. We use ITI to restore H$\alpha$ observations of scanned photographic film in the quality of the most recent CCD observations at KSO \cite{potzi2007scanning}. The film scans range from 1973 to 2000 and show degradations due to unequal illumination, atmospheric effects, low spatial resolution and digitalization. Figure \ref{figure:film_series}a shows samples across the full observing period. Similar to Sect. \ref{section:kso} we apply a center-to-limb correction, but use a resolution of 512$\times$512 pixels. This is inline with our objective of homogenizing the data series, where we primarily adjust the global appearance, while we neglect small scale details. We apply ITI to correct degradations and match the observations with the high-quality KSO data series.

In Fig. \ref{figure:film_series}b we show the reconstructed observations and a KSO observation from 2013, taken with the recent instrument. The ITI images appear similar to the recent KSO observation, leading to a more homogeneous observation series. We note that ITI also accounts for atmospheric degradations (observation from 1980-04-29) and that enhanced images appear similar to the KSO observations in Fig. \ref{figure:kso_combined}a. More detailed comparisons are given in the supplementary materials (App. \ref{section:detailed_film}).

\subsection{Approximation of observables -\newline STEREO/EUVI-to-SDO/HMI magnetograms}
\label{section:stereo_mag}

In case of missing observables, a reasonable approach is to estimate the missing information based on proxy data. The STEREO mission observes the Sun in four EUV channels, but has no magnetic field imager onboard. We use ITI to complement the STEREO observations, by generating LOS magnetograms based on STEREO EUV filtergrams. Solar far-side magnetograms with deep learning were first obtained by Kim \textit{et al.} \cite{kim2019solar}, by training a neural network for SDO data and applying it to STEREO observations. In this study, we directly translate STEREO EUV filtergrams to the SDO domain. Similar to Sect. \ref{section:stereo_soho_sdo} we use the EUV filtergrams of STEREO and SDO as low- and high-quality domain respectively. In addition, we add the SDO LOS magnetograms to the high-quality domain. Thus, the generator AB translates sets of images with four channels to sets of images with five channels. For the magnetograms we use the unsigned magnetic flux, in order to prevent unphysical assumptions about the magnetic field polarity, which would require a global understanding of the solar magnetic field (see App. \ref{section:sdo_euv_to_mag}).

For our training setup we use a discriminator model for each channel and single discriminator for the combined set of channels, this provides three optimization steps that enable the estimation of realistic magnetograms. (1) The single channel discriminator for the magnetograms assures that the generated magnetograms correspond to the domain of real magnetograms, independent of their content. (2) The content of the image is bound by the multi channel discriminator, that verifies that the generated magnetograms are consistent with EUV filtergrams of the SDO domain. (3) From the identity transformation B-B, we truncate the magnetogram from the SDO input data and enforce the reconstruction of the same magnetogram (see Sect.~\ref{section:training}).

\begin{figure}%
\includegraphics[width=\linewidth]{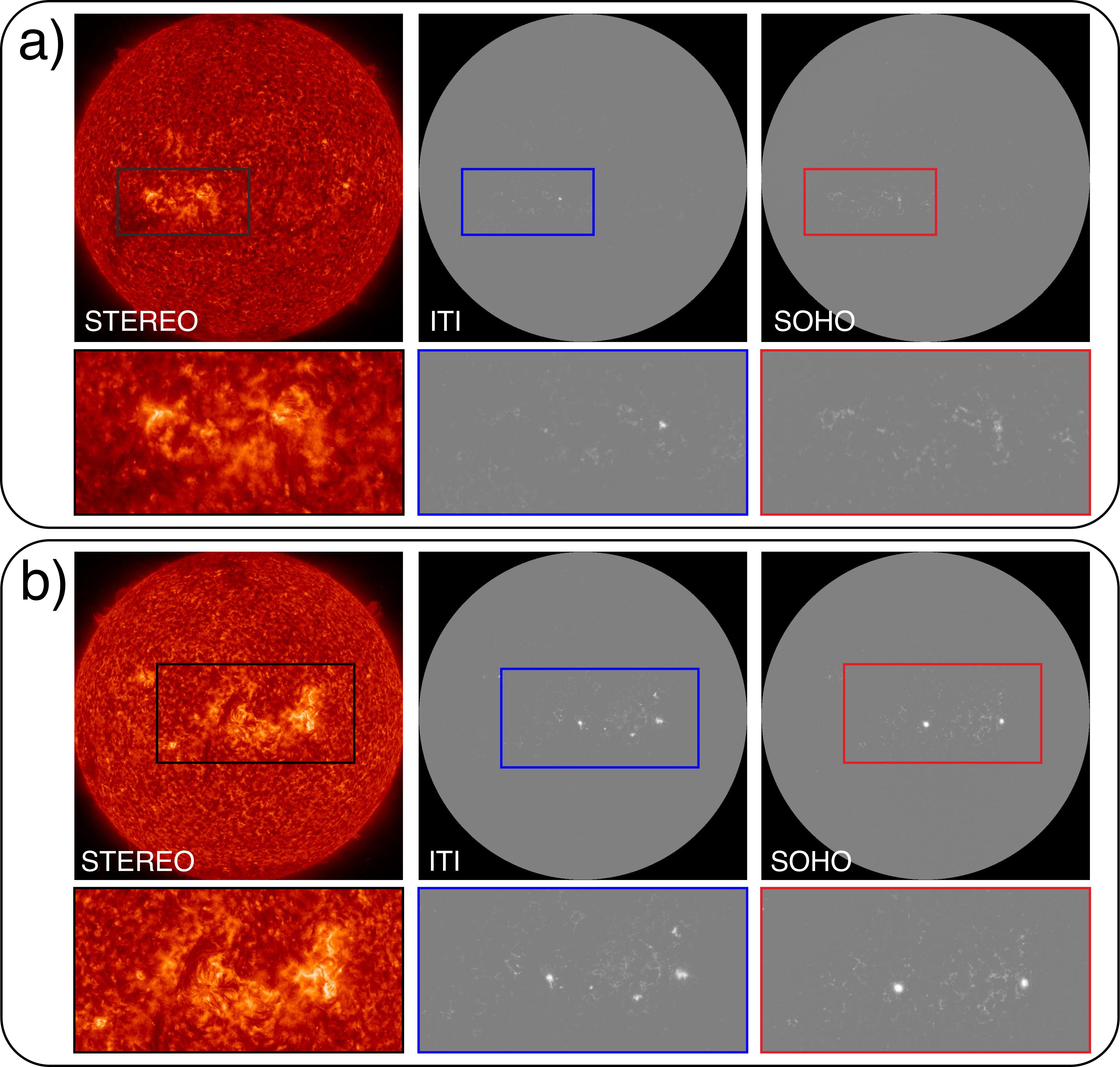}
\caption{Comparison between the synthetic ITI and overlapping SOHO/EIT magnetograms. In panel a and b we show observations from 2006-12-27 13:20 and 2007-01-07 19:20, respectively. STEREO/EUVI observations of the 304~$\AA$ channel are given for comparison and scaled by their maximum and minimum value. The LOS magnetograms show the absolute values of magnetic field strength and are scaled linearly between -2000 and 2000 Gauss. The flux distribution at the full-disk scale is in agreement between the ITI and SOHO/MDI observations. The comparison of the active region in a) shows that the ITI magnetogram matches the overall flux distribution. The active region in b) shows larger deviations that mainly originate from confined regions (i.e., sunspots) where ITI overestimates the magnetic field strength. An animated version is provided in the supplementary materials (Movie 3).}
\label{figure:stereo_sdo_mag}
\end{figure}

The different vantage points of STEREO and SDO do not allow for a direct comparison of estimated ITI and real SDO/HMI magnetograms, but partially overlapping observations were obtained by SOHO/MDI at the beginning of the STEREO mission (November 2006). Here, we use these SOHO/MDI magnetograms as reference to assess the validity of the estimated ITI magnetograms. Figure \ref{figure:stereo_sdo_mag} shows two examples of STEREO/EUVI 304 $\AA$ observations, the corresponding ITI magnetograms and the real SOHO/MDI reference magnetograms. We note a good agreement of the sunspot positions, that are not obvious from the EUV filtergrams. The magnetic elements appear similar in their distribution and magnitude, although we note a slight overestimation of magnetic field strength. The two examples show the variations of the generated magnetograms, where the magnetogram in panel a is close to the actual observation, while the magnetogram in panel b shows more deviations. The biggest differences originate from the overestimation of magnetic field strengths, which especially affects regions with strong magnetic flux, where extended magnetic elements can be falsely contracted to a sunspot (i.e., spurious sunspot in panel b). An analysis of the temporal stability and the accompanying movie are given in the supplementary material (App. \ref{section:consistency_stereo_mag}; Movie 3)

\begin{figure}%
\centering
\includegraphics[width=\linewidth]{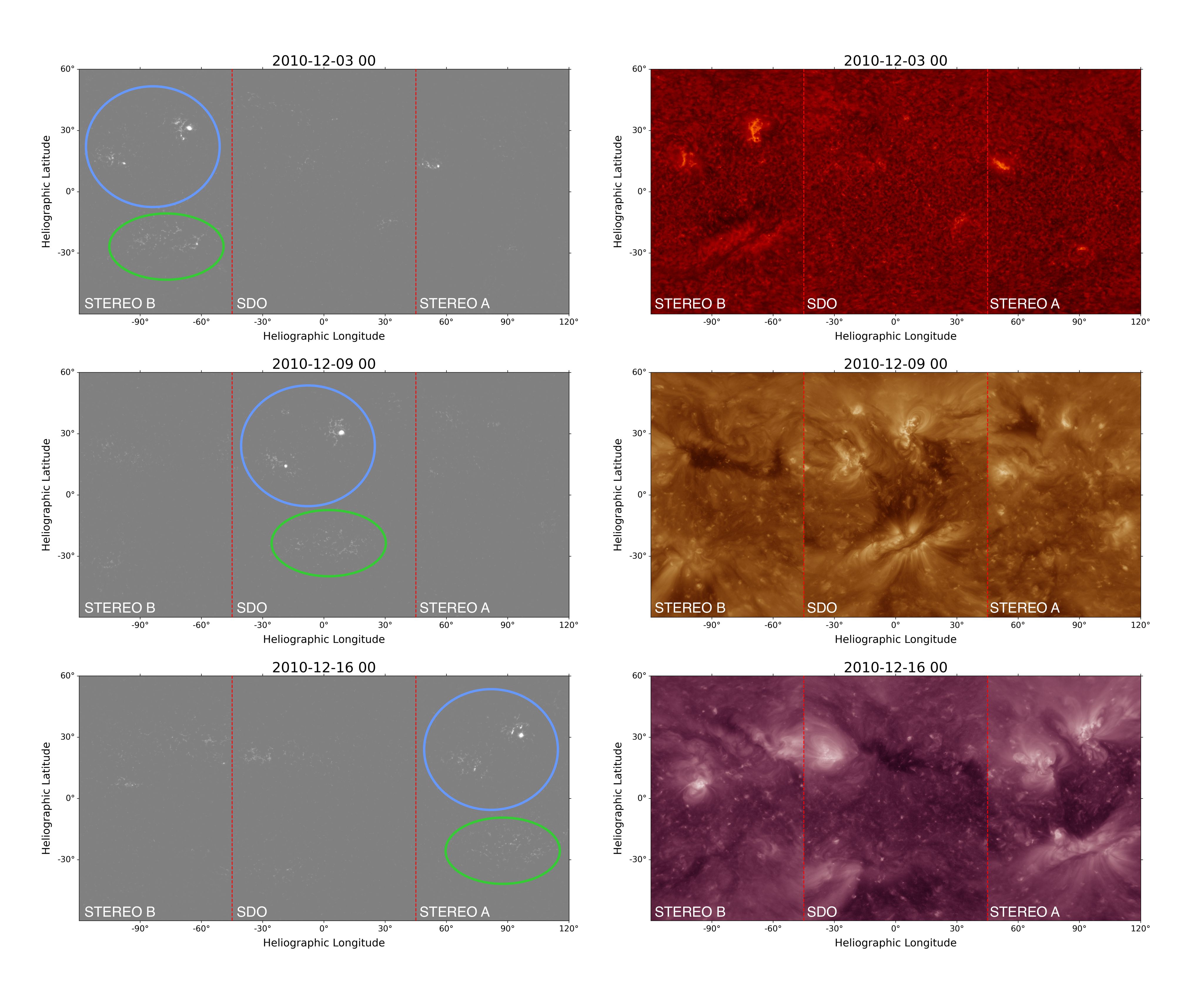}
\caption{Three consecutive observations of heliographic synchronic magnetograms (left) and reference EUV maps (right). We compare ITI magnetograms obtained from STEREO EUV filtergrams with real observations from SDO/HMI. We show the same active regions (blue circle) as observed by each instrument over 16 days. In the top row and bottom row the active regions are observed by STEREO B and STEREO A, respectively. The comparison to the real SDO/HMI observation (middle row) shows that ITI detects a similar magnetic flux distribution. Both ITI magnetograms show sunspot configurations with the characteristic tilt of preceding and following sunspot (Joy's law), but deviate from the SDO/HMI observation in terms of stronger and more confined magnetic fields (i.e., more sunspots). The green circle indicates the magnetic field configuration of the filament, that can be seen in the EUV observations. The ITI magnetograms show the expected topology of two opposing magnetic fields that sustain the filament channel and align with the SDO/HMI observation. An animated version of this figure is provided online (Movie 4). 
 ITI performs a domain translation that also includes the EUV channels. As can be seen from the heliographic EUV maps, the translated EUV channels smoothly integrate with the SDO observations.}
\label{figure:stereo_sdo}
\end{figure}

A direct application of the far-side magnetograms are full-disk heliographic synchronic maps of magnetic flux. We combine the ITI magnetograms from STEREO A and B together with the real SDO/HMI magnetograms into heliographic maps. On the left in Fig. \ref{figure:stereo_sdo} we show the magnetograms over 13 days, where the active regions (blue circle) are centered for STEREO~B, SDO and STEREO~A, respectively. On the right in Fig. \ref{figure:stereo_sdo}, we show the EUV filtergrams for comparison (c.f., Sect. \ref{section:stereo_soho_sdo}). The overall distribution of magnetic flux and the position of the major sunspots is consistent with the SDO observation, although we note that again small sunspots appear that are inconsistent with the SDO/HMI observation (panel a). Other magnetic features, such as the filament (green circle), can also be clearly identified in the ITI magnetograms. In all magnetograms the distinct separation between the two regions of strong magnetic flux is present, corresponding to the expected magnetic field topology.

The primary limitation to assess the validity of our approach is the availability of aligned observations. We employ the translation of SDO/AIA filtergrams to SDO/HMI magnetograms as reference task, which we can fully quantify. We identify that the main challenge is to estimate the exact position and the correct magnetic field strength of sunspots and pores (see App. \ref{section:sdo_euv_to_mag} for more details).

\subsection{Similarity of image distributions}
\label{section:image_distribution}

The Fréchet inception distance (FID) is a commonly used metric for GANs to estimate the quality of synthesized images and is a measure for the distance between two image distributions \cite{heusel2017frechet}. The qualitative samples in this section show an increase in perceptual quality. To verify that this improvement applies throughout the data set we use the FID to compare the similarity of the high-quality data sets to the corresponding low-quality data sets and their ITI enhanced version. The results in Table \ref{table:fid} show that for all instruments and channels the ITI enhanced images are closer to the high-quality image distribution than the low-quality distribution. This demonstrates that our method is able to map images closer to the high-quality distribution, or in other words, leads overall to a perceptually higher image quality.


\begin{table}
\caption{Evaluation of the FID for each application. We estimate the similarity of the ITI enhanced images and the original images to the reference high-quality observations. Lower values indicate a better perceptual image quality.}             
\label{table:fid}      
\centering                          
\begin{tabular}{| l || r r |}        
\hline
 & \multicolumn{2}{c|}{FID} \\
Instruments & original & ITI  \\    
\hline                        
   SDO/HMI $\xrightarrow{}$ Hinode continuum & 29.0 & \textbf{11.8}\\
   SOHO/EIT $\xrightarrow{}$ SDO/AIA 171 \AA & 38.2 & \textbf{4.1}\\
   SOHO/EIT $\xrightarrow{}$ SDO/AIA 193 \AA & 28.0 & \textbf{5.9}\\
   SOHO/EIT $\xrightarrow{}$ SDO/AIA 211 \AA & 2.9 & \textbf{1.0}\\
   SOHO/EIT $\xrightarrow{}$ SDO/AIA 304 \AA & 33.6 & \textbf{7.6}\\
   SOHO/MDI $\xrightarrow{}$ SDO/HMI magnetogram & 25.3 & \textbf{5.2}\\
   STEREO/EUVI $\xrightarrow{}$ SDO/AIA 171 \AA & 11.0 & \textbf{6.2}\\
   STEREO/EUVI $\xrightarrow{}$ SDO/AIA 193 \AA & 32.4 & \textbf{23.7}\\
   STEREO/EUVI $\xrightarrow{}$ SDO/AIA 211 \AA & 17.5 & \textbf{6.6}\\
   STEREO/EUVI $\xrightarrow{}$ SDO/AIA 304 \AA & 7.7 & \textbf{6.4}\\
   KSO LQ $\xrightarrow{}$ KSO HQ H$\alpha$ & 4.8 & \textbf{2.6}\\
   KSO Film $\xrightarrow{}$ KSO CCD H$\alpha$ & 26.0 & \textbf{8.1}\\
\hline                                   
\end{tabular}
\end{table} 

\section{Discussion}
\label{section:discussion}


In this study we presented a method for the homogenization of data sets, stable enhancement of solar images and restoration of long-term image time series. We applied unpaired image-to-image translation to use observations of the most recent instruments to enhance observations of lower quality. Our method does not require any spatial or temporal overlaps and can account for the translation between instruments with different field-of-view, resolution and calibration, which makes our method applicable to many astrophysical imaging data sets. The approach is purely data driven, which avoids assumptions about the mapping between high- and low-quality samples and improves the applicability to real observations. 

The use of paired domain translations between different instruments, typically requires extensive pre-processing, where observations need to be matched in time and aligned at the pixel-level. Both operations can typically not be perfectly satisfied and strongly reduce the available data for model training. Here, unpaired domain translation enables a new range of applications, without extensive pre-adjustments, and comparable performance to paired translations (c.f., Sect. \ref{section:sdo_euv_to_mag} and \cite{zhu2017unpaired}).

The five applications were selected to provide a comprehensive evaluation of the developed ITI method. We assessed 1) the pixel-wise matching of enhanced features, 2) the ability to intercalibrate multi-instrument data series, 3) the quality improvement as estimated by an independent quality assessment method, 4) the applicability to data sets that share no common observing periods, and 5) the validity and limitations of approximated observables. All samples presented in Sect. \ref{section:results} show a strong perceptual quality increase, reconstruction of faint details from a diverse set of degradations and are consistent across the full solar disk. The evaluation of the distance between the image distributions demonstrates quantitatively that this perceptual quality improvement applies throughout the full data set. Our method preserves measured intensity and flux values, such that the model results can be treated analogously to real observations.

We specifically addressed the question of the validity of the enhancement by comparing our results to real observations of the high-quality instruments. The enhanced images show throughout a strong similarity to the real observations and reveal a valid reconstruction of fibrils in the penumbra of sunspots (Fig. \ref{figure:hmi_hinode_comparison}), faint coronal loops (Fig. \ref{figure:soho_combined}b) and blurred plage regions (Fig. \ref{figure:kso_combined}a). This enhancement beyond the instrumental resolution, can be related to the inference of information from the high-quality image distribution. 


The concept of translating between image domains has also implications for multi-channel environments. We showed that with the simultaneous translation of image channels, and requiring the consistency of the combined set of channels, information about the image content can be exchanged between related channels (SOHO-to-SDO; Fig. \ref{figure:soho_combined}b) and missing channel information can be synthesized (far-side magnetograms; Sect. \ref{section:stereo_mag}). The restoration of H$\alpha$ film observations demonstrates that historic series can be homogenized with recent observation series, even if no overlapping observation periods exist (Sect. \ref{section:film}).

The quantitative evaluations demonstrate that our method achieves a stable performance across the full data sets. The pixel-wise comparison shows that our method outperforms a state-of-the-art deconvolution approach (SDO/HMI-to-Hinode continuum, Sect. \ref{section:hmi_hinode}), and quality assessment of ground-based observations (Sect. \ref{section:kso}) suggest that ITI images are close to the high-quality image domain. The analysis of the full-Sun EUV light-curves (Sect. \ref{section:stereo_soho_sdo}) suggests that our method also provides a calibration of photon counts. The comparison to a baseline calibration shows that this task is non-trivial, and that the feature-dependent translation of ITI leads to more consistent calibrated long-term series.

While the model can draw information from the context at larger and intermediate scales (e.g., separation of fibrils; Sect. \ref{section:hmi_hinode}), the structures at the smallest scales can not be reconstructed. This owes to the fact that the spatial information is degraded beyond the point of reconstruction, and the enhanced features solely match a common pattern of the high-quality distribution (e.g., small granules). Enhanced images can be used for a better visualization of blurred features or to mitigate degradations, but they should not be used for studying small scale features. The main application lies in the data calibration and assimilation, which allow us to study unified data sets of multiple instruments and to apply automated algorithms to larger data sets without further adjustments or preprocessing. 
Similarly, algorithms that perform feature detections or segmentations can profit from enhanced images (e.g., identification of magnetic elements, granulation tracking, filament detection).

The demonstrated applications are of direct use in solar physics. 1) The results of the HMI-to-Hinode translation allow for a continuous monitoring of the Sun in a resolution of 0.15~arcsec pixels, producing full-disk images with unprecedented resolution. As can be seen from Fig. \ref{figure:hmi_hinode_full_disk}, the high-resolution observations of Hinode/SOT provide only a partial view on the objects of interest. The enhanced HMI images can provide useful context information by accompanying high-resolution observations. This applies to both the spatial extent as well as the extent of the time series by additional observations before and after the Hinode/SOT series. 2) The homogeneous series of space-based EUV observations enables the joined use of the three satellite missions. The data set provides a resource for the study of solar cycle variations (cf. \cite{chatzistergos2019analysis, hamada2020new}), contributes additional samples for data-driven methods and enables the application of automated methods that were developed specifically for SDO/AIA data to the full EUV data series without further adjustments \cite{jarolim2021chronnos, armstrong2019fast}. 3) The correction of atmospheric effects in ground-based observations can be operated in real-time (approximately 0.44 seconds per observation on a single GPU). This allows to obtain more consistent observations and assists methods that are sensitive to image degradations (e.g., flare detection; \cite{veronig2016spaceweather}). 4) With the reconstruction of the H$\alpha$ film scans, we provide a homogeneous series starting in 1973, suitable for the study of solar cycle variations. 5) The generated STEREO far-side magnetograms give an estimate of the total magnetic flux distribution, which can provide a valuable input for space-weather awareness \cite{kim2019solar}. Information about the magnetic polarities is required for the further application to global magnetic field extrapolations \cite{jeong2020ai_mag_pfss}. From the patch-wise translation the inference of global magnetic field configurations (e.g., Hales law), would be arbitrary and was therefore omitted in this study. For all the considered applications, our trained model processes images at higher rates than the cadence of the instruments, allowing the application in real-time and fast reconstruction of large data sets.

The extension to new data sets requires the acquisition of a few thousand images of low- and high-quality, where an alignment is not required, and the training of a new model. The translation is performed for observations of the same type (e.g., LOS magnetogram) or in the same (or similar) wavelength band with similar temperature response (e.g., EUV), but with different image quality, either reduced by atmospheric conditions or by instrumental characteristics, such as spatial resolution. The data sets need to provide similar observations in terms of features and regions, in order to avoid translation biases \cite{cohen2018distribution}. The estimation of magnetic field information based on EUV filtergrams illustrates an example where this condition is not strictly required. Here the image translation is constrained by the multi-channel context information and the learned high-quality image distribution. 

We demonstrated that our neural network learns the characteristics of real high-quality observations, which provides an informed image enhancement, where the considered high-quality data set implies a desired upper limit of quality increase.

\section{Methods}
\label{section:methods}

 
Generative Adversarial Networks (GANs) have shown the ability to generate highly realistic synthetic images \cite{goodfellow2014gan, radford2015unsupervised, karras2019stylegan2}. The training is performed in a competitive setup of two neural networks, the generator and discriminator. Given a set of images, the generator maps a random input from a prior distribution (latent space) into the considered image domain, while the discriminator is trained to distinguish between real images and synthetic images of the generator. The generator is trained to compete against the discriminator by producing images that are classified by the discriminator as real images. The iterative step-wise optimization of both networks allows the generator to synthesize realistic images and the discriminator to identify deviations from the real image distribution \cite{goodfellow2014gan}. By replacing the prior distribution with an input image, a conditional mapping between image domains can be achieved \cite{wang2018pix2pixhd, isola2017image}.

We use a GAN to generate highly realistic low-quality observations that show a large variation of degrading effects. We propose an informed image enhancement which uses domain specific knowledge to infer missing information. For this task, we employ a GAN to model the high-quality image distribution and constrain enhanced images to correspond to the same domain. With the use of a sufficiently large data set we expect that we can learn to correctly model the true image distributions and find a mapping that is applicable for real observations.


\subsection{Model training}
\label{section:training}

The primary aim of our method is to transform images from a given low-quality domain to a target high-quality domain. We refer to the high-quality domain as B and to the low-quality domain as A. In order to achieve an image enhancement that can account for various image degradations we aim at synthesizing realistic images of domain A based on images of domain B. The pairs of high-quality and synthetic low-quality images are used to learn an image enhancement. Thus, the training process involves mappings from A to B (A-B), as well as mappings from B to A (B-A).

The model setup involves four neural networks, two generators and two discriminators. The generators learn a mapping between A-B and B-A. The discriminators are used to distinguish between synthetic and real images of domain A and B. The training cycle for image enhancement uses a high-quality image (B) as input, which is translated by the generator BA to domain A. The synthetic degraded image is then restored by the generator AB (Fig.~\ref{figure:iti_BAB}). We optimize the generators to minimize the distance between the original and reconstructed image, as estimated by the reconstruction loss (cycle consistency). The simplest solution for this setting would be an identity mapping by both generators. We counteract this behavior with the use of discriminator A, which is trained to distinguish real images of domain A from synthetic images of generator BA. With this we constrain the generator BA to generate images that correspond to domain A and the generator AB to restore the original image from the degraded version. With the synthesis of more realistic low-quality images, we expect the generator AB to perform equally well for real low-quality images.

Similarly to Wang \textit{et al.} \cite{wang2018pix2pixhd} each discriminator network is composed of three individual networks, where we decrease the image resolution by a factor 1, 2, 4 for the three networks, respectively. With this the perceptual quality optimization is performed at multiple resolution levels, thus estimating small-scale features as well as more global relations.

We follow the training setup of Zhu \textit{et al.} \cite{zhu2017unpaired} and use three additional training cycles. The second cycle enhances the perceptual quality of high-quality images with the use of discriminator B. The mapping is again performed under the cycle consistency criteria. Thus, we start with a low-quality image and perform an A-B mapping, followed by a B-A mapping (Supplementary Fig.~\ref{figure:iti_ABA}). The additional training with discriminator B ensures that images produced by generator AB correspond to domain B, which adds an additional constrain for image enhancement and improves the perceptual quality. The last two training cycles ensure a consistency under identity transformations. To this aim, we translate images of domain A with the generator BA and images of domain B with the generator AB, and then minimize the reconstruction loss between the original and transformed images (Supplementary Fig.~\ref{figure:iti_id}). For differences in resolution we use bilinear upsampling and average pooling for downsampling.

By only using an image as input to our generator, the results are deterministic \cite{almahairi2018augmented}. In the present case we are interested in modeling various degrading effects and explicitly want to avoid the generation of synthetic noise, based on image features (e.g., solar features, instrumental characteristics). This task is often addressed as multimodal translation \cite{zhu2017multimodel, huang2018multimodal} and also relates to style transfer between images \cite{huang2017adain, karras2019stylegan2, choi2020stargan}. Here, we add an additional noise term to our generator BA, so that multiple degraded versions can be generated from a single high-quality image. For the generator AB we assume that there exists a unique high-quality solution for a given low-quality observation.

The cycle consistency of low-quality images is ensured by first estimating the noise of the original low-quality image. For this task, we employ an additional neural network which we term noise-estimator. We use the noise-estimator for the A-B-A and A-A mapping and randomly sample a noise term for the B-A-B mapping from a uniform distribution [0, 1]. The advantage of this approach is two fold. 1) The mappings A-B-A and A-A are not ambiguous, which allows for a clear separation of noise from high-quality images. 2) The explicit encoding of low-quality features into the noise term representation benefits the relation between the generated low-quality features and the noise term, by enforcing the use of the noise term in the generator \cite{choi2020stargan}. For both the A-B-A and A-A mapping we minimize the distance between the estimated noise of the original image and the estimated noise of the reconstructed image (Supplementary Fig.~\ref{figure:iti_id}; Sect.~\ref{section:iti_reconstruction}). This approach relates to image style transfer (e.g., \cite{johnson2016perceptual, karras2019stylegan2, choi2020stargan}), where we consider the low-quality features as style that we transfer to the high-quality images. 

Image quality can be addressed in terms of perceptual (e.g., mean-squared-error) and distortion (e.g., Wasserstein distance) quality metrics. The spanned space by both metrics is referred to as perception-distortion-plane, where the minimum on both metrics is not attainable \cite{blau2018perceptionDistortion}. In order to obtain the best quality, image enhancement algorithms should consider both metrics for optimization. This can be seen from image translation tasks, where the additional use of an adversarial loss achieved significant improvements in terms of image quality \cite{wang2018pix2pixhd, isola2017image}. We employ content loss, which shows a better correspondence to image features than pixel based metrics (e.g. MSE), as primary distortion metric and use adversarial loss of GANs for the perceptual optimization. From this setup we aim at achieving an optimal minimum in the perception-distortion plane.

Most full-disk observations exceed the computational limit when used in their full resolution. This can be overcome by reducing the resolution of the image or by training the neural network with image patches. In this study, we use image patches, in order to provide images with the highest resolution attainable. After model training, the neural network can be applied to full-resolution images, which is in most cases computationally feasible.

As described in Cohen \textit{et al.} \cite{cohen2018distribution}, the use of data sets with an unequal distribution of features is likely to produce artifacts. For this reason we balance during training between image patches with solar features (e.g., active regions, filaments) and quiet Sun regions. During training we randomly sample patches from a biased distribution. For Hinode/SOT we additionally consider the solar limb, such that patches are equally sampled across the full solar disk.

\subsection{Multi-channel architecture}
\label{section:channel_discriminator}

In the case of multiple image channels (e.g., simultaneously recorded filtergrams at different wavelengths) we employ multiple discriminator networks (Supplementary Fig. \ref{figure:multi_channel}). We use an individual discriminator for each of the image channels (\textit{single-channel}) and an additional discriminator that takes the combined channels as input (\textit{multi-channel}). Here, each discriminator represents again a set of three single discriminators at three different resolution scales (cf. Sect. \ref{section:training}). From the usage of the single-channel discriminators we expect a better matching of the individual channel domains without influences of the other channels. The multi-channel discriminator is capable to assess the validity between the channels (e.g., appearance of features across the channels). This concept is especially important for the estimation of observables (Sect. \ref{section:stereo_mag}), where the single-channel discriminator solely addresses the perceptual quality of the estimated observable, while the multi-channel discriminator restricts the estimated observable to be valid within the context of the other channels. 

We note that the multi-channel architecture has no strict mapping requirements (e.g., same filters for each channel), such that the number of input and output channels is flexible. Thus, additional observables can be approximated (e.g., more output channels than input channels) and a conversion between similar channels can be achieved (e.g., SOHO/EIT 195~$\AA$ to SDO/AIA 193~$\AA$). 

\subsection{Reconstruction loss}
\label{section:iti_reconstruction}

The cycle consistency serves as distortion metric for image enhancement. Pixel-based metrics (e.g., mean-absolute-error)
\begin{equation}
    \label{equation:iti_mae}
    \mathcal{L}_{MAE,BAB} = \mathbb{E} [\Vert x_B - G_{AB}(G_{BA}(x_B, z)) \Vert_1],
\end{equation}
can prevent large divergences between the original image $x_{B}$ and the reconstruction $G_{AB}(G_{BA}(x_B, z))$ but small shifts can cause a large increase of the reconstruction loss, which might provide a poor estimate of the image quality \cite{blau2018perceptionDistortion}. For this reason, we utilize in addition content loss, that compares feature similarity over pixel-wise similarity. Layer-wise activation of a neural network resemble extracted features and are therefore related to the content of the image. The content loss metric is computed by taking the mean-absolute-error (MAE) between the feature activation of the generated and original image. We define the content loss based on the discriminator, similar to Wang \textit{et al.} \cite{wang2018pix2pixhd} as
\begin{equation}
    \label{equation:iti_content}
    \mathcal{L}_{Content,BAB,j} = \mathbb{E} \sum_{i=1}^{4}\frac{1}{N_i}[\Vert D^{(i)}_{B,j}(x_B) - D^{(i)}_{B,j}(G_{AB}(G_{BA}(x_B, z)))\Vert_1],
\end{equation}
where $x_B$ refers to a high-quality image; $z$ to the noise term; $D^{(i)}_{B,j}$ to the activation layer $i$ of network $j$ from discriminator B; $G_{AB}$ and $G_{BA}$ to the generator AB and generator BA, respectively; and $N_i$ to the total number of features of activation layer $i$. For each of our discriminators we use all intermediate activation layers (cf. \cite{jarolim2020image}).

For consistency of the estimated noise terms, we minimize the MAE between the estimated noise term of the original low-quality image and its reconstructed version (A-B-A)
\begin{equation}
    \mathcal{L}_{Noise,ABA}= \mathbb{E} [\Vert NE(x_A) - NE(G_{BA}(G_{AB}(x_A), NE(x_A))) \Vert_1],
\end{equation}
and similarly for the identity mapping (A-A)
\begin{equation}
    \mathcal{L}_{Noise,AA}= \mathbb{E} [\Vert \textit{NE}(x_A) - NE(G_{BA}(x_A, \textit{NE}(x_A))) \Vert_1],
\end{equation}
where $\textit{NE}$ refers to the noise-estimator. To avoid the collapse of the noise term to a constant value, we minimize the MAE between the randomly sampled noise term and the estimated noise of the synthetic low-quality image (B-A-B)
\begin{equation}
    \mathcal{L}_{Noise,BAB}= \mathbb{E} [\Vert z - \textit{NE}(G_{BA}(x_B, z)) \Vert_1].
\end{equation}

Equations \ref{equation:iti_mae} and \ref{equation:iti_content} refer to the B-A-B cycle. The other training cycles are computed analogously, where we use the discriminator of the corresponding domain for the extraction of activation features (i.e., $D_A$ for images of domain A). For the total reconstruction loss $\mathcal{L}_{Rec.}$ we combine the MAE loss and content loss
\begin{equation}
\label{equation:iti_reconstruction}
\begin{aligned}
    \mathcal{L}_{Rec.} = {}
    &\lambda_{mae} \cdot (\mathcal{L}_{MAE,BAB} + \mathcal{L}_{MAE,ABA}) + \\
    &\lambda_{mae,id} \cdot (\mathcal{L}_{MAE,BB} + \mathcal{L}_{MAE,AA}) + \\
    &\lambda_{content} \cdot \frac{1}{4 M}\sum_{j=1}^{M} \left(\mathcal{L}_{Content,BAB,j} + \mathcal{L}_{Content,ABA,j}\right) + \\
    &\lambda_{content, id} \cdot \frac{1}{4 M}\sum_{j=1}^{M} \left(\mathcal{L}_{Content,BB,j} +  \mathcal{L}_{Content,AA,j}\right) + \\
    &\lambda_{noise} \cdot (\mathcal{L}_{Noise,ABA} + \mathcal{L}_{Noise,AA} + \mathcal{L}_{Noise,BAB}),
\end{aligned}
\end{equation}
where $M$ refers to the number of networks used for the discriminators and the $\lambda$ parameters are used to scale the individual losses. The factor $4 M$ is introduced to normalize the content loss for the number of discriminator layers per domain (cf. Eq. \ref{equation:iti_content}).


\subsection{Adversarial loss}
\label{section:iti_adversarial}

As originally proposed by Goodfellow \textit{et al.} \cite{goodfellow2014gan}, GANs are composed of a generating network (\textit{generator}) that produces a synthetic image from a random input vector (\textit{latent space}), and a discriminating network (\textit{discriminator}) that distinguishes between generated and real images. The training is performed in a competitive setup between the generator and discriminator. We optimize the discriminator $D$ and generator $G$ for the objective proposed by Mao \textit{et al.} \cite{mao2017least} (Least-Squares GAN). The discriminator A enforces the generation of low-quality images in the B-A-B cycle
\begin{equation}
    \label{equation:iti_adversarial_discriminator}
    \mathcal{L}_{D_{A,j}} = \mathbb{E}\left[(D_{A,j}(x_A) - 1)^2 \right] + \mathbb{E}\left[D_{A,j}(G_{BA}(x_B, z))^2\right],
\end{equation}
where $D_{A, j}$ refers to the $j$th network of discriminator A, $x_A$ to a real low-quality image, $z$ to the noise term and $G_{BA}(x_B, z)$ to a generated low-quality image by generator BA. The generator is optimized to achieve a classification as real image by each discriminator network
\begin{equation}
    \label{equation:iti_adversarial_generator}
    \mathcal{L}_{G,BA} = \mathbb{E}\left[ \frac{1}{M}\sum_{j=1}^{M} (D_{A,j}(G_{BA}(x_B, z)) - 1)^2\right].
\end{equation}
The loss for discriminator B and generator AB is computed in the same way, where the discriminator enforces the generation of high-quality images in the A-B-A cycle.

\subsection{Diversity loss}
\label{section:iti_diversity}

The variation of synthetic images is a frequently addressed topic for GANs \cite{gulrajani2017wgangp, radford2015unsupervised, choi2020stargan}. For conditional image translation a variation of generated images can be obtained by inducing additional noise terms \cite{huang2018multimodal, almahairi2018augmented} or by including style information \cite{huang2017adain, choi2020stargan}. Instance normalization achieved significant improvements in neural style transfer \cite{ulyanov2016instance}. The normalization is applied after each convolution layer by normalizing each feature channel for zero mean and unit variance
\begin{equation}
\label{equation:norm}
    \text{IN}(x) = \gamma \frac{x- \mu(x)}{\sigma(x)} + \beta,
\end{equation}
where $\mu(x)$ refers to the mean and $\sigma(x)$ to the variance across the spatial dimensions and $\gamma$ and $\beta$ to learnable parameters.

The effect of instance normalization can be interpreted as a normalization of the feature space in terms of style, where the affine parameters ($\gamma$ and $\beta$) act as a projection to a different style \cite{huang2017adain}. In the present case, the translation between different instruments can be understood as style transfer, where we transfer the high-quality style to images of low-quality. The affine parameters allow the network to learn a single style. For the training with image patches, we use running estimates of the normalization variables ($\mu$, $\sigma$) with a momentum of 0.01 (statistical parameters are computed according to $x_{new} = (1 - 0.01) \cdot x_{old}  + 0.01 \cdot x_{observed}$). 

For the generator BA we enable the generation of various low-quality images from a single high-quality image by including a noise term, which we sample from a uniform distribution [0, 1]. A frequent problem of GANs is mode collapse, where the network generates the same image independent of the noise term \cite{radford2015unsupervised, gulrajani2017wgangp}. We introduce an additional loss term to prevent mode collapse and increase the diversity of generated low-quality images \cite{yang2019diversity, choi2020stargan, mao2019mode}. We sample two independent noise terms ($z$ and $\hat{z}$) for the same high-quality image $x_B$ and compute the content loss between the resulting two images
\begin{equation}
\begin{split}
    \mathcal{L}_{Content,Diversity,j}(x_B, z, \hat{z}) = \sum_{i=1}^{4}\frac{1}{N_i}[\Vert D^{(i)}_{A,j}(G_{BA}(x_B, z)) -\\  D^{(i)}_{A,j}(G_{BA}(x_B, \hat{z}))\Vert_1],
\end{split}
\end{equation}
where $G_{BA}$ refers to the generator BA (cf. Eq. \ref{equation:iti_content}). We scale the difference in content loss by the distance of the noise terms and apply the logarithm to the result, which leads to an increased loss for nearly identical images and reduces divergences for large differences
\begin{equation}
    \label{equation:iti_diversity}
    \mathcal{L}_{Diversity} = \mathbb{E} \left[ \log{\frac{\frac{1}{4 M}\sum_{j=1}^{M} \mathcal{L}_{Content,Diversity,j}(x_B, z, \hat{z})}{\Vert z - \hat{z} \Vert_1}} \right].
\end{equation}

\subsection{Combined loss}
\label{section:iti_combined_loss}

The training cycles for the generators and discriminators are performed end-to-end, where we alternate between generator and discriminator updates. Our full generator objective is given by combining Eq. \ref{equation:iti_reconstruction}, \ref{equation:iti_adversarial_generator} and \ref{equation:iti_diversity}
\begin{equation}
\begin{split}
    \min_{G_{AB}, G_{BA}, NE} (\mathcal{L}_{Reconstruction} - \lambda_{divsersity} \cdot \mathcal{L}_{Diversity} + \\
    \lambda_{adversarial} \cdot \left( \mathcal{L}_{G,BA} + \mathcal{L}_{G,AB} \right) ),
\end{split}
\end{equation}
where the $\lambda$ parameters are used to scale the individual losses. The discriminator objective is obtained from Eq. \ref{equation:iti_adversarial_discriminator}:
\begin{equation}
    \min_{D_A, D_B} \left( \frac{1}{M}\sum_{j=1}^{M} (\mathcal{L}_{D_{A,j}} + \mathcal{L}_{D_{B,j}})  \right).
\end{equation}
$D_A$ and $D_B$ refer to the combined discriminators for images of domain A and B, respectively ($D_A = \{D_{A,1}, ..., D_{A,M}\}$).

\subsection{Data set}
\label{section:iti_data_set_pairs}

Supplementary Table \ref{table:iti_data_set} summarizes the considered instruments, observing wavelengths, the amount of data samples and the number of independent patches that can be extracted. The images are normalized such that saturations are avoided, which we found beneficial for model training. Further details on data pre-processing, correction of device degradation and normalization is given in Appendix~\ref{section:preprocessing}. We apply a strict temporal separation of our train and test set, where we consider the last two months of each year only for model evaluations, and use a buffer of one month between training and test set to avoid a potential memorization \cite{liu2021reliability}. In general, we always consider February to September for our training set, while observations from November and December correspond to the test set. In addition, we exclude the SDO/AIA+HMI observations from 2010 that overlap with the SOHO/EIT+MDI observations from training.

The observations from SDO, STEREO and SOHO are taken at high cadence. For each data set we randomly sample observations from the full mission lifetime (SDO: 2010-2020, \url{http://jsoc.stanford.edu}; STEREO: 2006-2020, Solar Data Analysis Center via \url{http://virtualsolar.org}; SOHO: 1996-2011, STEREO Science Center via \url{http://virtualsolar.org}). For Hinode/SOT observations we use 2$\times$2 binned red continuum observations at 6684 $\AA$ of the SOT Broadband Filter Imager (BFI) as high-quality reference. In the time between 2007 and 2016 we select a single observation from each observation series taken with BFI (\url{https://darts.isas.jaxa.jp/solar/hinode}). For the KSO H$\alpha$, we utilize the image quality assessment method by Jarolim \textit{et al.} \cite{jarolim2020image}, to automatically assemble a data set of ground-based KSO observations that suffer from atmospheric degradations (e.g., clouds, seeing), in the time between 2008 and 2019. As high-quality reference we use the synoptic archive of contrast enhanced KSO observations, that comprises manually selected observations of each observing day. In addition we perform a quality check and remove all observations of reduced image quality (\url{http://cesar.kso.ac.at}). The scanned KSO film H$\alpha$ filtergrams include all observations taken in the time period from 1973 to 2000, independent of image quality (\url{ftp://ftp.kso.ac.at/HaFilm/}). We manually remove all observations that suffer from severe degradations (e.g., missing segments, strong overexposure, misalignment). Observations that suffer from minor degradations (e.g., spurious illumination, clouds, artifacts), are still considered for image enhancement.

With the use of unpaired data sets there is typically no limitation in data samples. For all our applications we select a few thousand images and extend them in case of diverging adversarial training.

\subsection{Training Parameters}

For our model training we use the Adam optimizer with a learning rate of 0.0001 and set $\beta_1 = 0.5$ and $\beta_2 = 0.9$ \cite{kingma2014adam}. As default value we set all $\lambda$ parameters to 1. The content loss is scaled by 10 and the identity losses are scaled by 0.1 (s.t. $\lambda_{content} = 10$, $\lambda_{content,id} = 1$, $\lambda_{mae,id} = 0.1$). For all our models we track running statistics of the instance normalization layers ($\gamma$, $\beta$ in Eq. \ref{equation:norm}) for the first 100 000 iterations and fix the values for the remaining iterations. For all space-based observations we set $\lambda_{diversity} = 0$, such that the main focus is the adaption of the instrumental characteristics and the generation of spurious artifacts is neglected (e.g., pixel errors). The training is performed with batch size 1 and the size of image patches is chosen based on our computational capabilities.

In the Supplementary Table \ref{table:training} we summarize the parameters of our model training. For each run we stop the training in case of convergence (no further improvement over 20 000 iterations). For the more complex low-quality features and limited number of samples, the generator BA can diverge to unrealistic low-quality images at later iterations. Here, we reduce the training iterations, where we note a stable training.

\subsection{Alignment of HMI and Hinode images}
\label{section:align_hmi}

For comparing continuum observations from SDO/HMI and Hinode/SOT we perform a pixel-wise alignment of image patches. We only consider images that contain solar features (e.g., we avoid quiet-Sun and limb regions) and that have been recorded within $<$10 seconds. To this aim, we scale the Hinode observations to 0.15 arcsecs and the HMI observations to 0.6 arcsecs per pixel. We apply our ITI method to upsample the HMI images by a factor of 4 in order to match the Hinode observations.
From the header information of the Hinode files we select the corresponding HMI region. We co-register the translated HMI patches with the Hinode patches using the method by \cite{reddy1996fftregister}, that determines translational shifts and rotational differences, based on phase-correlation of the Fourier-transformed images. After the co-alignment we truncate the edges (80 pixels per side) to avoid boundary artifacts. 
As baseline image enhancement method we apply a Richardson-Lucy deconvolution \cite{richardson192deconv}, using the estimated point-spread-function of HMI by Yeo \textit{et al.} \cite{yeo2014hmi_psf} ($K_1$), and applying 30 iterations. The deconvolved images are bilinear upsampled by a factor 4 and aligned analogously to the ITI image patches.

To compare the counts between Hinode and HMI-deconvolved images we analogously extract patches from the training set and compute the mean ($\overline{I}$) and standard deviation ($\sigma$) over the full data, which is used to adjust the HMI intensities as
\begin{equation}
\label{equation:calibration}
    I_{calibrated} = (I_{A} - \overline{I_{A}}) \cdot \frac{\sigma_{B}}{\sigma_{A}} + \overline{I_{B}},
\end{equation}
where $A$ refers to the source calibration (HMI) and $B$ to the target calibration (Hinode).

Typically there is a difference in the calibration between the reference and enhanced image, even after corrections, that leads to a varying shift in the metrics. For our evaluation we primarily focus on the difference of structural appearance, where we first normalize each image patch based on the maximum and minimum value.

\subsection{Calibration of EUV data sets}
\label{section:euv_baseline}

As calibration baseline we assume that the quiet-Sun is invariant to solar-cycle variations. We extract quiet-Sun regions by selecting all on-disk pixels that are below the estimated threshold. For this, we compute the mean and standard deviation per channel from observations sampled across the timeline and set the threshold to mean $+ 1 \sigma$. For each data set, we compute the average mean and standard deviation of the quiet-sun pixels. For our baseline calibration we adjust the mean and standard deviation between the channels of the individual instruments using Eq. \ref{equation:calibration}.

\subsection{Computation of the FID}
\label{section:fid_computation}

For the computation of the FID, we convert the test sets of each dataset to gray-scale images. Differences in resolution between high- and low-quality data sets, are adjusted by bilinear upsampling. The reconstructed STEREO magnetograms are not included in this evaluation, since there exists no low-quality data set for comparison (Sect. \ref{section:stereo_mag}).
We use an IncepctionV3 model with the weights of the tensorflow implementation for the FID. The FID is evaluated for each application and channel separately, where we use the full resolution images (i.e., no resizing) and do not normalize the inputs.

\section{Code availability}

The codes and trained models are publicly available: 
\url{https://github.com/RobertJaro/InstrumentToInstrument}

The software is designed as a general framework that can be used for automatic image translation and for training similar applications. Data preprocessing and download routines are provided.

\section{Data availability}

We publicly provide the enhanced and filtered KSO film observations with a resolution of 512$\times$512 pixels.
For the other presented data sets we provide lists of files used for training and evaluation (publicly available). The provided models can be used to reproduce the data sets from our evaluation. 

\bibliographystyle{naturemag}
\bibliography{references}

\section{Acknowledgements}
This research has received financial support from the European Union’s Horizon 2020 research and innovation program under grant agreement No. 824135 (SOLARNET).
The authors acknowledge the use of the Skoltech HPC cluster Zhores and Arkuda for obtaining the results presented in this paper \cite{zacharov2019zhores}.
This research has made use of SunPy v3.0.0 \cite{sunpy_software2020}, an open-source and free community-developed solar data analysis Python package \cite{sunpy_community2020}.
The calculation of the FID was performed with the codes by \cite{Seitzer2020FID}.

\section{Author contributions}

R.J. developed the method and led the writing of the paper, A.V. contributed to the conceptualization of the study and writing of the paper, W.P. contributed to the KSO data analysis, T.P. contributed to the HPC computations. All authors discussed the results and commented on the manuscript.

\newpage

\renewcommand{\figurename}{Supplementary Figure}
\setcounter{figure}{0}

\renewcommand{\tablename}{Supplementary Table}
\setcounter{table}{0}

\begin{appendix}


\section{Supplementary Figures}

\begin{figure}[!htb]%
\includegraphics[width=\linewidth]{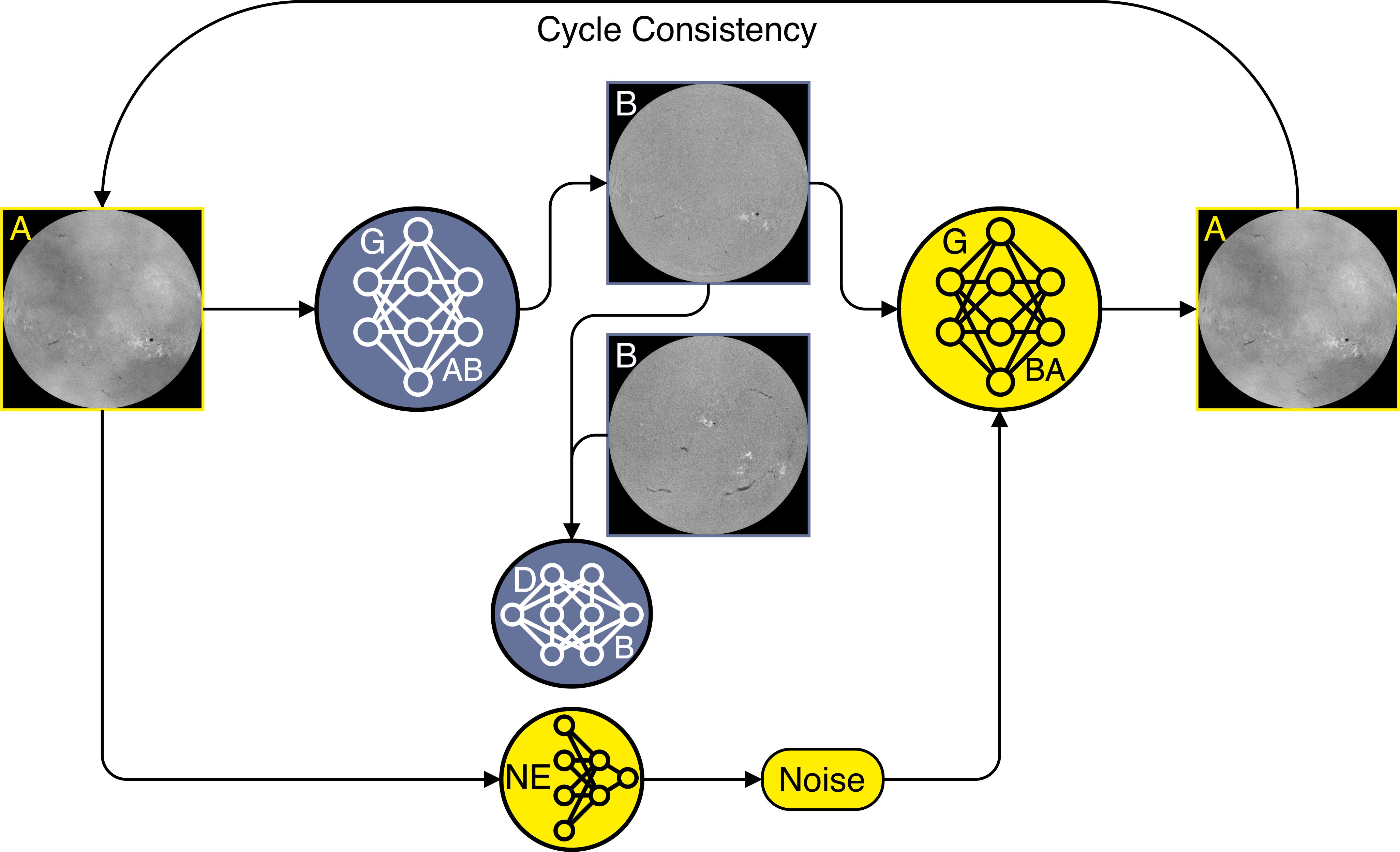}
\caption{Model training for image enhancement. Images of domain A are translated to domain B by generator AB (blue). The enhanced images are translated back to domain A by generator BA (yellow). The perceptual image quality is optimized with the use of discriminator B, which distinguishes between real images of domain B (bottom) and enhanced images by generator AB (top). The noise term for the reconstruction of the original images by generator BA is obtained by the noise-estimator (NE).}
\label{figure:iti_ABA}
\end{figure}

\newpage

\begin{figure}[!htb]%
\includegraphics[width=\linewidth]{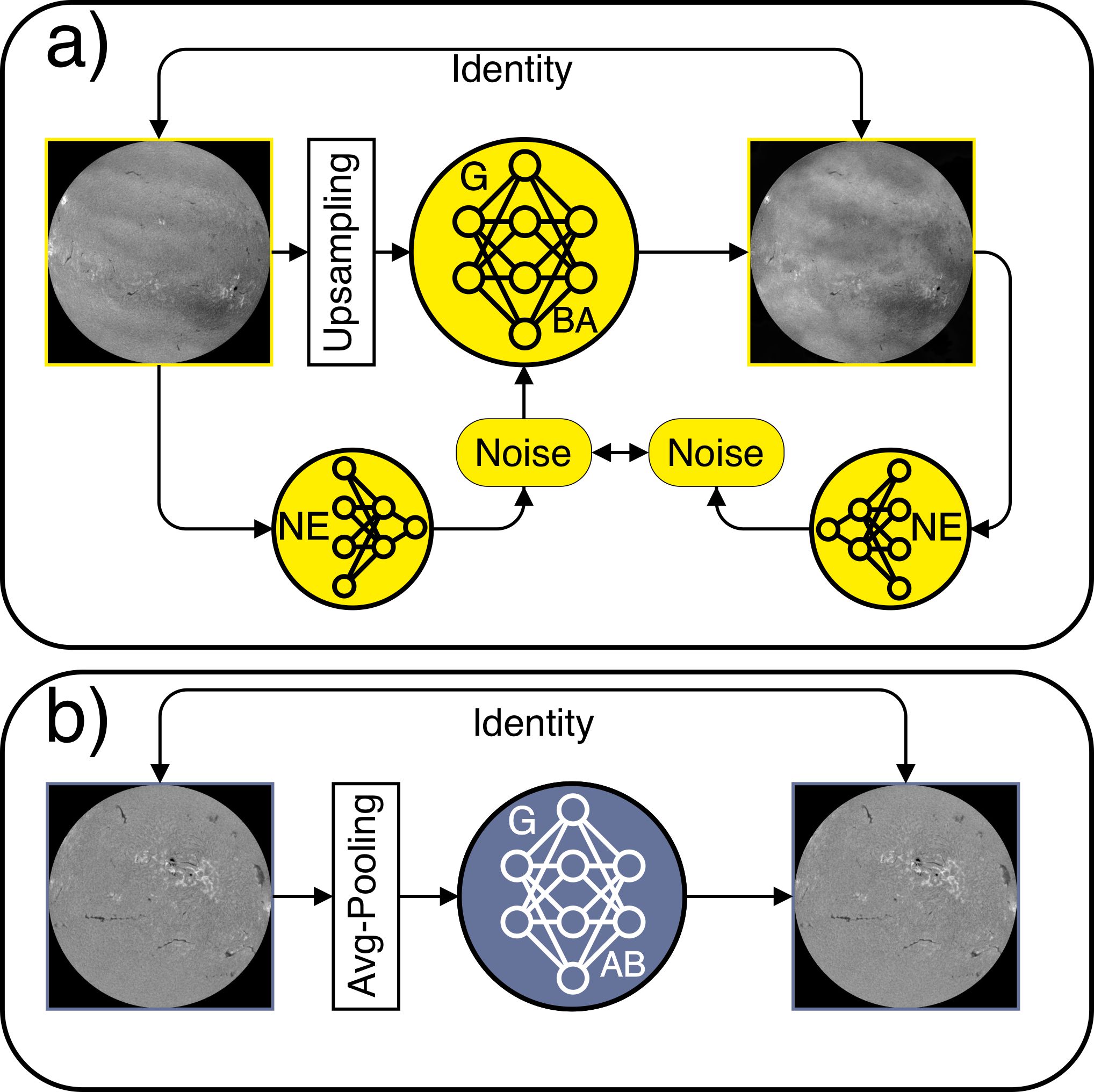}
\caption{Model training for the identity transformations. Images are optimized to remain unchanged for a mapping into the same domain. Top panel: mapping from domain A to domain A by generator BA (yellow). Bottom panel: mapping from domain B to domain B by generator AB (blue). The noise term for generator BA is obtained from the original image by the noise-estimator (NE). In addition the network is optimized to estimate the same noise before and after transformation (top panel). In case of different resolutions the low-quality images are first upsampled for A-A and downsampled for B-B.}
\label{figure:iti_id}
\end{figure}

\begin{figure}[!htb]%
\includegraphics[width=0.8\linewidth]{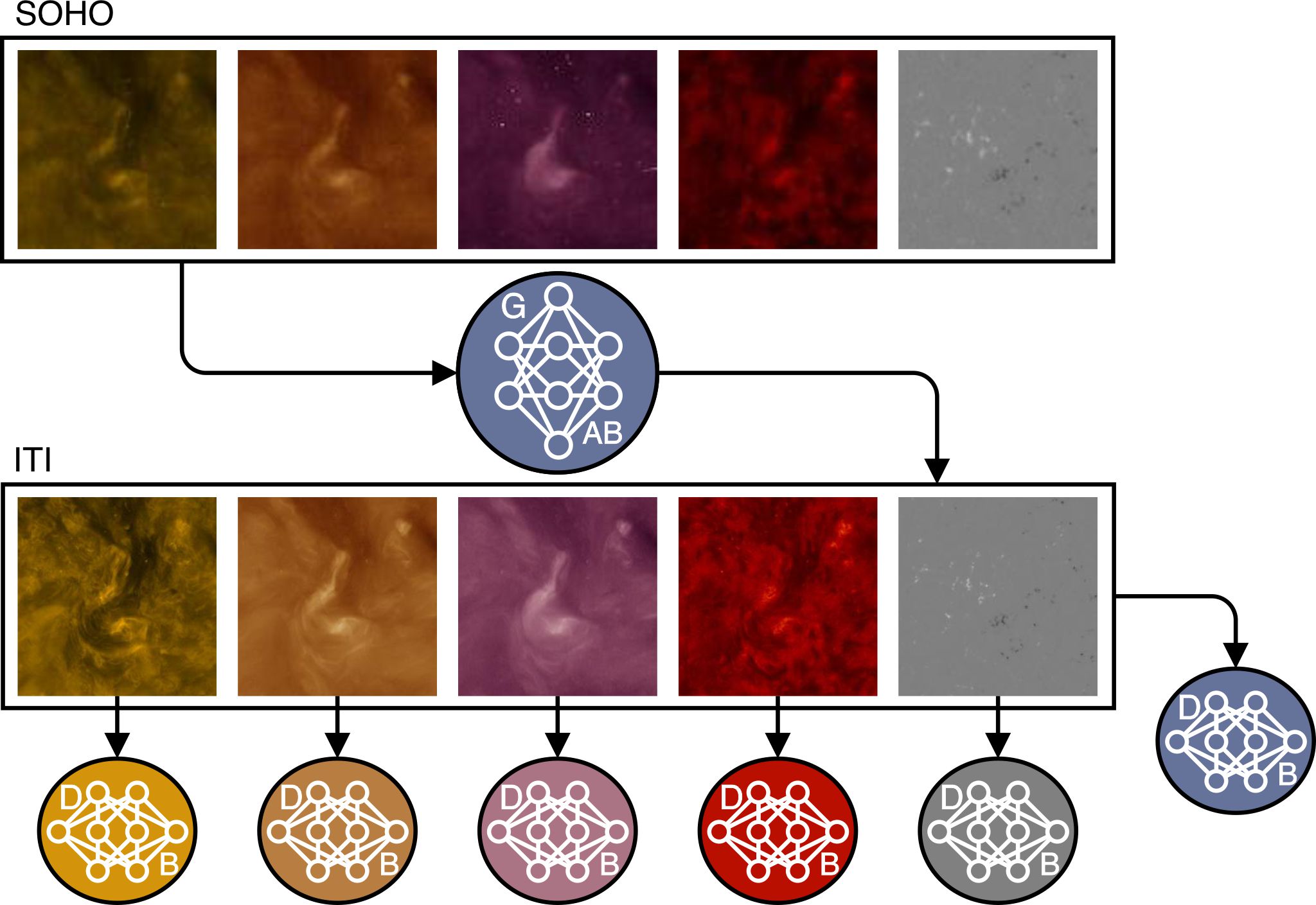}
\caption{Example of the multi-channel use of the discriminator. Each image channel of the source low-quality domain (top) is simultaneously translated to the target high-quality domain (bottom) by the generator AB (SOHO/EIT+MDI-to-SOHO/AIA+HMI). For each translated image channel a separate discriminator is used. The inter-channel consistency is assessed by a single discriminator that uses the combined set of channels as input.}
\label{figure:multi_channel}
\end{figure}

\newpage

\section{Supplementary Tables}

\begin{table*}[!h]
\caption{Data sets for training of Instrument-To-Instrument translation. The effective number of samples refers to the number of independent patches that can be extracted from the training set. Observations of Hinode/SOT have a varying field-of-views, the provided value gives the upper limit. }             
\label{table:iti_data_set}      
\centering                          
\begin{tabular}{|c || l r r r|}        
\hline                 
Instrument & Observables & Train & Effective Number of Samples & Test \\    
\hline                        
   SDO & 171, 193, 211, 304, LOS magnetogram & 9562 & 3.8e4/15.3e4 & 2404\\
   SOHO & 171, 195, 284, 304, LOS magnetogram & 4855 &  31.1e4 & 1410\\
   STEREO & 171, 195, 284, 304 & 3146 & 20.1e4 & 822\\
   SDO/HMI & continuum & 2751 & 2.8e4 & 595\\
   Hinode/SOT & continuum & 2643 & $\sim$5286 & 668\\
   KSO synoptic & H$\alpha$ & 2106 & 3.4e4 & 422\\
   KSO LQ & H$\alpha$ & 1461 & 23.4e4 & 384\\
   KSO Film & H$\alpha$ & 4067 & 1.6e4 & 772\\
\hline                                   
\end{tabular}
\end{table*}

\begin{table*}[!h]
\caption{Overview of the Training parameters. For each instrument translation we train our model for the given number of iterations (Iter.). The upsampling parameter gives the number of upsampling operations that are applied to the low-quality images (Up). The diversity parameter is only considered for ground-based observations ($\lambda_{d}$). The patch size is given for the low-quality samples, such that the high-quality patch size corresponds to (Patch Size)$\cdot 2^{\text{Up}}$.}             
\label{table:training}      
\centering                          
\begin{tabular}{| l || c c c c |}        
\hline
Instruments & Iter. & Up & $\lambda_{d}$ & Patch Size [pixels]  \\    
\hline                        
   SDO/HMI $\xrightarrow{}$ Hinode continuum & 260 000 & 2 & 0 & 160$\times$160 \\
   SOHO/EIT+MDI $\xrightarrow{}$ SDO/AIA+HMI & 460 000 & 1 & 0 & 128$\times$128 \\
   STEREO/EUVI $\xrightarrow{}$ SDO/AIA & 520 000 & 2 & 0 & 128$\times$128 \\
   KSO LQ $\xrightarrow{}$ KSO HQ H$\alpha$ & 260 000 & 0 & 1 & 256$\times$256 \\
   KSO Film $\xrightarrow{}$ KSO CCD H$\alpha$ & 240 000 & 0 & 1 & 256$\times$256 \\
   STEREO/EUVI $\xrightarrow{}$ SDO/AIA+HMI & 340 000 & 0 & 0 & 256$\times$256 \\
\hline                                   
\end{tabular}
\end{table*}

\begin{table*}
\caption{Evaluation of the intercalibration of SOHO, STEREO and SDO observations. We compare our ITI calibration with the baseline calibration, where we consider smoothed light-curves for STEREO to mitigate the different vantage points to SDO. For SOHO we compare temporally aligned observations. We compute the mean absolute difference (MAE) between the calibrated and SDO light-curves and the cross-correlation (CC) between the calibrated STEREO and SDO light-curves.}             
\label{table:euv_calibration}      
\centering                          
\begin{tabular}{| l l || c c |}        
\hline
    Instrument & Method & MAE $\downarrow$ & CC $\uparrow$  \\    
\hline                        
    STEREO/EUVI - 171 \AA & baseline & 18.6 & 0.68 \\
     & ITI & 12.7 & 0.76 \\
    \hline
    STEREO/EUVI - 195 \AA & baseline & 29.0 & 0.97 \\
     & ITI & 17.5 & 0.97 \\
    \hline
    STEREO/EUVI - 284 \AA & baseline & 20.1 & 0.96 \\
     & ITI & 10.2 & 0.97 \\
    \hline
    STEREO/EUVI - 304 \AA & baseline & 2.1 & 0.91 \\
     & ITI & 2.2 & 0.92 \\
    \hline
    STEREO/EUVI - avg & baseline & 17.4 & 0.88 \\
     & ITI & 10.7 & 0.90 \\
    \hline
    \hline
    SOHO/EIT - 171 \AA & baseline & 7.7 & - \\
    & ITI & 12.4 & - \\
    \hline
    SOHO/EIT - 195 \AA & baseline & 9.2 & - \\
    & ITI & 8.5 & - \\
    \hline
    SOHO/EIT - 284 \AA & baseline & 3.3 & - \\
    & ITI & 1.8 & - \\
    \hline
    SOHO/EIT - 304 \AA & baseline & 1.3 & - \\
    & ITI & 2.4 & - \\
    \hline
    SOHO/EIT - avg & baseline & 5.4 & - \\
    & ITI & 6.3 & - \\
\hline                                   
\end{tabular}
\end{table*}

\section{Instruments and Data}
\label{section:data_set}

This section describes the characteristics of the considered instruments for the evaluation of our method. The pair-wise correspondence of the data sets is discussed in Sect.~\ref{section:iti_data_set_pairs}.

The Solar Dynamics Observatory (SDO) is a space-based mission located in a circular geosynchronous orbit, that provides science data since 1 May 2010 \cite{pesnell2012sdo}. For our study we use EUV filtergrams from the Atmospheric Imaging Assembly (AIA; \cite{lemen2012aia}), and line-of-sight (LOS) magnetograms and continuum observations from the Helioseismic and Magnetic Imager (HMI; \cite{schou2012hmi}). Both instruments provide solar full-disc images with a spatial sampling of 0.6~arcsec pixels and a spatial resolution of 1.5~arcsec. The data are recorded by 4096$\times$4096 pixel CCDs. The AIA instrument operates at a cadence of 12~seconds and provides EUV filtgrams in seven band passes. In this study, we consider Fe IX (171 $\AA$), Fe XII, XXIV (193 $\AA$), Fe XIV (211 $\AA$) and He II (304 $\AA$) filtergrams. The emission lines are primarily formed in the solar corona and chromosphere and are associated with peak temperatures of ion formation at $6.3\cdot10^5$~K (Fe IX; quiet corona, upper transition region), $1.6\cdot10^6$~K, $2.0\cdot10^7$~K (Fe XII, XXIV; corona, hot flare plasma), $2.0\cdot10^6$~K (Fe XIV; active-region corona) and $5.0\cdot10^4$~K (He II; chromosphere, transition region) \cite{lemen2012aia}. The HMI instrument provides maps of the photospheric magnetic field strength and orientation by observing polarizations of the Fe I absorption line (6173~$\AA$). The continuum observations are calculated from six points in the Fe I line  \cite{schou2012hmi}. For both the HMI magnetograms and continuum observations, we use the 720 seconds series.

The Solar and Heliospheric Observatory (SOHO) is a space-based mission, located at Lagrange point L1, that was launched in December 1995 \cite{domingo1995soho}. In this study we use data from the Extreme-ultraviolet Imaging Telescope (EIT; \cite{delaboudiniere1995eit}) and LOS magnetograms from the Michelson Doppler Imager (MDI; \cite{scherrer1995mdi}). Both instruments provide full-disc 1024$\times$1024 pixels images of the Sun at a spatial sampling of 2.6 arcsec pixels and about 5~arcsec spatial resolution. From EIT we use Fe IX (171~$\AA$), Fe XII (195~$\AA$), Fe XV (284~$\AA$) and He II (304~$\AA$) filtergrams. The peak temperatures for ion formation are $1.3\cdot10^6$~K, $1.6\cdot10^6$~K, $2.0\cdot10^6$~K and $8.0\cdot10^4$~K, respectively \cite{delaboudiniere1995eit}. The MDI instrument is the predecessor of SDO/HMI and derives the LOS magnetograms in the Ni I 6768 A absorption line \cite{scherrer1995mdi}.

The Solar Terrestrial Relations Observatory (STEREO) is a twin-satellite mission that operates two identical satellites on two orbits close to 1~AU. The orbits are selected to lead to a yearly separation of the spacecrafts by about 45 degree, as viewed from the Sun \cite{kaiser2008stereo}. The mission was launched in 2006 and provides stereoscopic observations of the Sun since then.  In this study, we use the Exteme Ulraviolet Imager (EUVI; \cite{wulser2004euvi}) of the Sun–Earth Connection Coronal and Heliospheric Investigation (SECCHI; \cite{howard2008secchi}) instrument. The imager provides full-disc filtergrams with 2048$\times$2048 pixels with a spatial sampling of about 1.6~arcsec pixels and 3.2~arcsec spatial resolution \cite{howard2008secchi}. Similarly to SOHO/EIT, filtergrams of Fe IX (171~$\AA$), Fe XII (195~$\AA$), Fe XV (284~$\AA$) and He II (304~$\AA$) are recorded. The associated peak temperatures are in the same range as for the SOHO/EIT filters (see above).

The Solar Optical Telescope (SOT) onboard the Hinode satellite (launched in 2006) is a 50~cm aperture telescope that provides high-resolution images of partial fields of the Sun \cite{tsuneta2008sot}. The instrument comprises multiple broad- and narrow-band filters and provides spatial resolutions of up to 0.2 arcsec and a pixel scale of 0.054~arcsec~pixels. In this study, we use continuum observations centered at a wavelength of 6684~$\AA$ of the Broadband Filter Imager (BFI). The BFI instrument provides a field-of-view of 218"$\times$109", producing images with 4096$\times$2048 pixels, and operates at a cadence $<10$s \cite{tsuneta2008sot}. Observations are recorded on demand.

Kanzelh\"ohe Observatory for Solar and Environmental Research (KSO; \url{https://kso.ac.at/}) provides ground-based solar full-disk H$\alpha$ filtergrams at a spatial resolution of 2 arcsec. The data are recorded by a 2048$\times$2048 pixel CCD corresponding to a sampling of about 1 arcsec pixels. KSO regularly takes H$\alpha$ images at a cadence of 6 seconds and provides a fully automated data reduction and data provision pipe line, which allows for data access in near real time \cite{otruba2003halphaKSO, potzi2015real, potzi2021kso}. The current instrument setup is in operation since 2008. The H$\alpha$ line is formed by absorption at 6563 $\AA$ in the solar chromosphere and by cooler plasma in the solar corona (filaments).

At KSO, regular monitoring of the Sun in the H$\alpha$ spectral line is provided since 1973. The observations during 1973 and 2000 were recorded on photographic film at a cadence of about 4 minutes and were later digitized in 1024$\times$1024 pixels format \cite{potzi2007scanning, potzi2008scanningResults}.

\section{Data pre-processing}
\label{section:preprocessing}

\subsection{Space-based observations}

For each observation we center the Sun, normalize the field-of-view to 1.1 solar radii and crop the frame, such that the extent of the solar-disk is independent of yearly variations caused by the elliptic orbit (cf. \cite{galvez2019machine}). We correct for instrumental degradation and normalize the exposure time \cite{boerner2014photometric}. For STEREO/EUVI, SOHO/EIT, Hinode/SOT we use the IDL eit\_prep, secchi\_prep and fg\_prep routines, respectively. For SDO/AIA we use the SDO autocalibration by \cite{santos2021aiacalibration}. The STEREO/EUVI instrument shows significant degradations of the 304~$\AA$ channel, that are not considered by the STEREO prep routine ($\sim 50 \%$). We correct for this degradation by extracting the quiet-Sun regions over the full data set and fitting the degradation as a first-order polynomial. The maximum intensity value of each instrument channel is estimated from the maximum intensity values over the full data set by $\hat{I}_{max} = \text{mean}(I_{max}) + 0.5 \cdot \text{std}(I_{max})$, where $I_{max}$ refers to the maximum intensity values of the images and $\text{std}$ to the standard deviation. We clip negative values and values larger than the estimated maximum intensity and normalize to [0, 1]. We apply a asinh stretch for EUV filtergrams and scale the data afterwards to the interval [-1, 1], such that it corresponds to a $\tanh$ activation function
\begin{equation}
    \hat{x} = \frac{\text{asinh}(x/a)}{\text{asinh}(1/a)} 2 - 1,
\end{equation}
with $a=0.005$. This stretching function provides for a logarithmic behavior for large values and a linear behavior for small values. For LOS magnetograms we use a linear normalization of values between $[-1000, 1000]$ Gauss to $[-1, 1]$ and set all off-limb values to zero. 

For the continuum observations of SDO/HMI and Hinode we analogously estimate the maximum value and scale linearly between 0 and the maximum value.

\subsection{Ground-based observations}
For all KSO observations we apply the same preprocessing. We shift the Sun to the image center, crop the field-of-view to 1 solar radius and resize it to 512$\times$512 pixels and 1024$\times$1024 pixels for the Film-to-CCD and Low-to-High translation, respectively. The center-to-limb variation is corrected by plotting the theoretical correction $\mu = \cos{\theta}$ against the intensity values of the image, where $\theta$ refers to the heliocentric angle, and fitting a fourth order polynomial that gives the intensity correction $I_{corr}(\mu)$  (cf. \cite{diercke2018counter}). For KSO observations of the recent instrument (CCD), we scale values between 0.65 and 1.5 to $[-1, 1]$ with a asinh stretch $(a=0.5)$. For the KSO film observations we apply the same scaling but adjust the scaling to values between 0.39 and 1.94 to account for the variations of the degraded observations. We apply the preprocessing per frame and set all off-limb pixels to $-1$. 

\section{Model architecture}
\label{section:iti_architecture}

\begin{figure}%
\includegraphics[width=\linewidth]{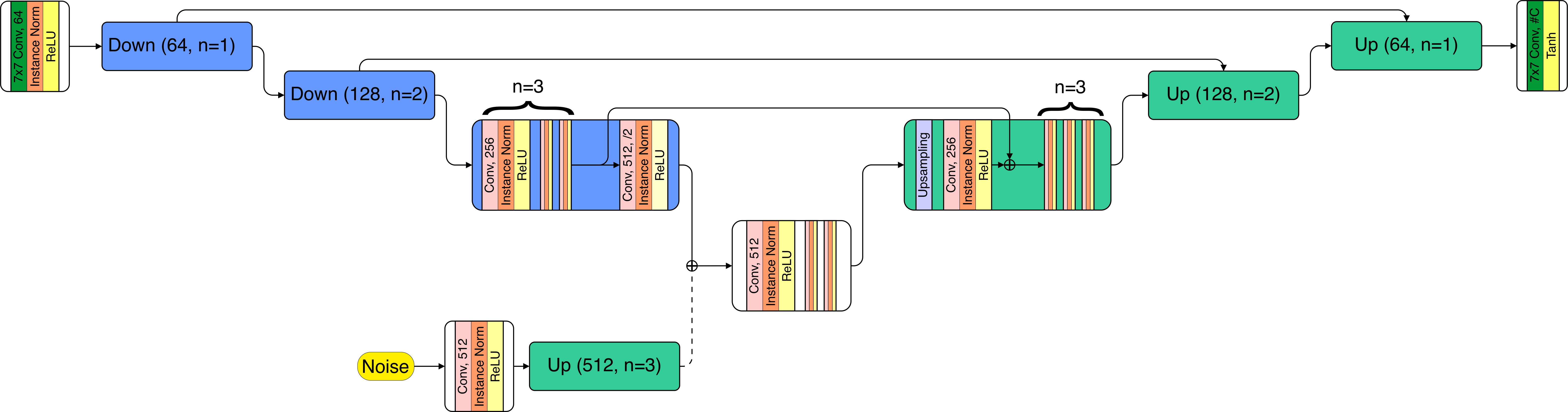}
\caption{Overview of the generator architecture. Our generators are composed of an input block, three downsampling blocks, an intermediate block, three upsampling blocks and an output block. The input block transforms the image to 64 filter channels. Each downsampling block reduce the spatial resolution by 4 while increasing the number of filter channels by 2. The number of convolutional blocks is linearly increased. The upsampling blocks operate in reverse order. At the intermediate layer the noise term for generator BA is added by concatenation (dashed line). The noise term is matched to the dimensions of the intermediate layers by an initial convolution block and an upsampling block. Features from the downsampling blocks are concatenated to the upsampling block features at the same resolution. The output image is obtained by a convolution layer followed by a $\tanh$ activation.}
\label{figure:iti_generator}
\end{figure}

For image translation tasks the concept of multi-scale architectures has shown great success. Skip connections have demonstrated the ability to enhance training performance and to preserve spatial details \cite{he2016deep, ronneberger2015unet}.

For our generators we employ convolution blocks similar to Karras \textit{et al.} \cite{karras2017progressive} and introduce skip connections at each resolution level, similar to Ronneberger \textit{et al.} \cite{ronneberger2015unet}. The full overview of our generator models is given in Supplementary Fig.~\ref{figure:iti_generator}. We use convolutional blocks composed of a convolutional layer with a kernel size of 3$\times$3, followed by an instance normalization \cite{ulyanov2016instance} and a ReLU activation. In order to retain the image dimensions and to avoid boundary artifacts, we use reflection padding before the convolution layers \cite{wang2018pix2pixhd}. Our network is composed of three downsampling blocks, an intermediate block, followed by three upsampling blocks. A downsampling block consists of $n$ convolutional blocks followed by a convolutional block with stride 2 (downsampling by 2). While downsampling we double the number of filter channels. For an upsampling block we first apply bilinear upsampling, reduce the number of filter channels by a factor of 2 with a convolutional block and apply $n$ additional convolutional blocks. At each resolution level we use skip connections between the downsampling and upsampling blocks. Hereby the features prior to downsampling are concatenated with the features after the first convolutional block in the upsampling block. The input image is transformed by an initial convolutional block  with a kernel size of 7$\times$7 and 64 filter channels. The three downsampling blocks are consecutively applied with $n=1,2,3$. The intermediate layer consists of three convolution blocks with the same number of filters as the last downsampling block (512). The upsampling blocks are applied in the inverse order ($n=3,2,1$). The output image is obtained by a convolution layer with kernel size 7$\times$7, followed by a $\tanh$ activation function.

For the generator BA we include the noise term by concatenation with the features of the last downsampling block (prior to the intermediate block). The noise term is transformed by a convolutional block with 512 filter channels and matched to the dimensions of the last downsampling block by an upsampling block with $n=3$ and 512 filter channels (dashed line in Supplementary Fig.~\ref{figure:iti_generator}).

Our networks are fully convolutional, therefore the training can be performed with image patches, while the evaluation is performed with full resolution images. For the noise term we use 16 channels and the spatial dimensions are adjusted to the considered image size ($1/16$ of the input image resolution). The instance normalization accounts for the global style transfer (see Sect.~\ref{section:iti_diversity}), while the noise term accounts for localized degradations in the image.

For the discriminators we use the architecture introduced in Wang \textit{et al.} \cite{wang2018pix2pixhd}, where each discriminator is composed of three individual networks that operate on different scales. The discriminators are composed of four consecutive stride 2 convolutions with instance normalization and ReLU activation. We start with 64 filter channels and consecutively increase them by a factor 2 for each layer. The discriminators output is obtained by a convolution layer with one filter channel. Therefore, each discriminator provides a grid of predictions instead of a single output \cite{isola2017image}. For the noise estimator we use the same architecture as for the discriminators, adjust the number of filters of the last layer to the noise dimensions (16) and apply a final sigmoid activation function.

For an image enhancement that involves a resolution increase, we extend the generator AB by additional upsampling blocks and the generator BA by additional downsampling blocks. In the case of images with multiple channels (e.g., multiple filtergrams, magnetograms), we translate all channels simultaneously. For each image channel we use an additional discriminator that ensures the correct representation of the generated channel, independent of the other image channels. The primary discriminator is adjusted to take all filter channels into account simultaneously to ensure the consistency between the channels. For an unequal number of channels between domain A and B, we adjust the identity cycles by truncating additional channels and extending missing channels with zeros.


\subsection{Ablation study}

We perform an ablation study to identify the most important components of our network architecture. The primary difference to existing domain translation models are the introduced skip-connections, the running statistics of the Instance-Normalization, the one-sided noise factor and the channel-wise discriminators.  We found that the InstanceNorm with running statistical values of the normalization parameters is required to obtain an equal translation across the full images. The application of a sample-wise normalization would lead to a translation dependent on the selected patch, while we aim to train with small patches and then extend our model to full-resolution images for inference. The noise term is considered necessary to provide a consistent mapping between the image domains (one-to-many mapping).

In Table \ref{table:ablation} we vary our model configuration in terms of number of filters (i.e., model capacity), depth of the generator (i.e., field-of-view), skip-connections, and discriminator mode. We refer to 'multi' when we use an additional discriminator for each image channel, and to 'single' if we only employ a discriminator for the combined set of channels. Each model is trained for 180,000 iterations for the SOHO-To-SDO translation task (Sect. \ref{section:stereo_soho_sdo}). We evaluate each model in terms of the FID averaged over all channels. We associate a lower FID with a stronger similarity to the reference high-quality image distribution, and consequently a better performance. The results show that the skip-connections provide throughout a better performance, which we associate with a better spatial relation to the input image and a more efficient gradient update. Moreover, the models without skip-connections are more unstable for deeper architectures, as can be seen from the diverged training of the depth 4 setting. An insufficient capacity leads to poor results (e.g., 32 filters), while our configurations with 64 filters show the best results. From the unpaired data set, our training is typically not limited in terms of data, with the implication that more parameters lead to a better performance. The comparison of the discriminator modes shows that the additional channel-wise discriminators as additional training target significantly improve the overall performance.

The improvement for the depth 4 configuration is only marginal, while the model is three times larger. For this reason, the version with 64 filters, depth 3 and skip-connections was considered for the presented domain translations in Sect. \ref{section:results}.

\begin{table}
\caption{Ablation study of our model configuration. We vary our architecture in dependence of number of filters, generator depth, skip-connections and discriminator mode. The number of parameters refer to the generator AB. We estimate the performance in terms of mean FID (lower better).}             
\label{table:ablation}      
\centering                          
\begin{tabular}{| c c c c c || c |}        
\hline
Filters & Depth & Skip-Connections & Discriminator & Parameters [M] & FID  \\    
\hline                        
   32 & 3 & \xmark & multi & 3.62 & 15.6 \\
   32 & 4 & \xmark & multi & 14.65 & 22.8 \\
   64 & 3 & \xmark & multi & 14.44 & 13.6 \\
   64 & 4 & \xmark & multi & 58.55 & - \\
   32 & 3 & \cmark & multi & 3.81 & 9.9 \\
   32 & 4 & \cmark & multi & 15.44 & 8.5 \\
   64 & 3 & \cmark & multi & 15.22 & 7.5 \\
   64 & 4 & \cmark & multi & 61.68 & 6.8 \\
   64 & 3 & \cmark & single & 15.22 & 14.6 \\
   64 & 4 & \cmark & single & 61.68 & 11.2 \\
\hline                                   
\end{tabular}
\end{table}

\section{Supplementary Evaluation}

\subsection{Extended comparison to image deconvolution}
\label{section:app_hmi_hinode}

For the evaluation of enhanced SDO/HMI continuum observations, we match active region patches and compute the peak signal-to-noise ratio (PSNR), and the structural similarity index measure (SSIM). In addition we evaluate the PSNR prior to normalizing the images and the FID of this subset (Supplementary Table \ref{table:hmi_comparison}). As can be seen from the calibrated HMI images without the deconvolution, the offset in calibration dominates the quality metric. For this reason, the images need to be normalized to assess the content similarity. ITI shows a worse calibration than the baseline, which owes to the fact that the baseline is fitted solely with data similar to the test set, while ITI is trained to enhance full-disk observations.  In Supplementary Fig. \ref{figure:hmi_hinode_comparison} we give additional samples of the aligned SDO/HMI, ITI and Hinode/SOT continuum observations, as well as the corresponding difference maps.

\begin{table}
\caption{Evaluation of HMI-to-Hinode enhanced observations. We compare our ITI method, calibrated HMI images, and additionally deconvolved images with aligned Hinode observations. We evaluate the PSNR, SSIM, and FID for the normalized images. The unnormalized PSNR uses image patches without a prior-normalization. For the full data set we compute the distance to the Hinode image distribution (FID).}             
\label{table:hmi_comparison}      
\centering                          
\begin{tabular}{| l || c c c c |}        
\hline
    Method & PSNR $\uparrow$ & unnormalized PSNR $\uparrow$ & SSIM $\uparrow$ & FID $\downarrow$  \\    
\hline                        
    HMI - calibrated & 18.4 & 24.7 & 0.55 &  74.1 \\
    HMI - deconvolved & 23.3 & 24.4 & 0.59 &  78.7 \\
    ITI & 24.7 & 20.3 & 0.64 &  44.6 \\
\hline                                   
\end{tabular}
\end{table}

\begin{figure}%
\centering
\includegraphics[width=\linewidth]{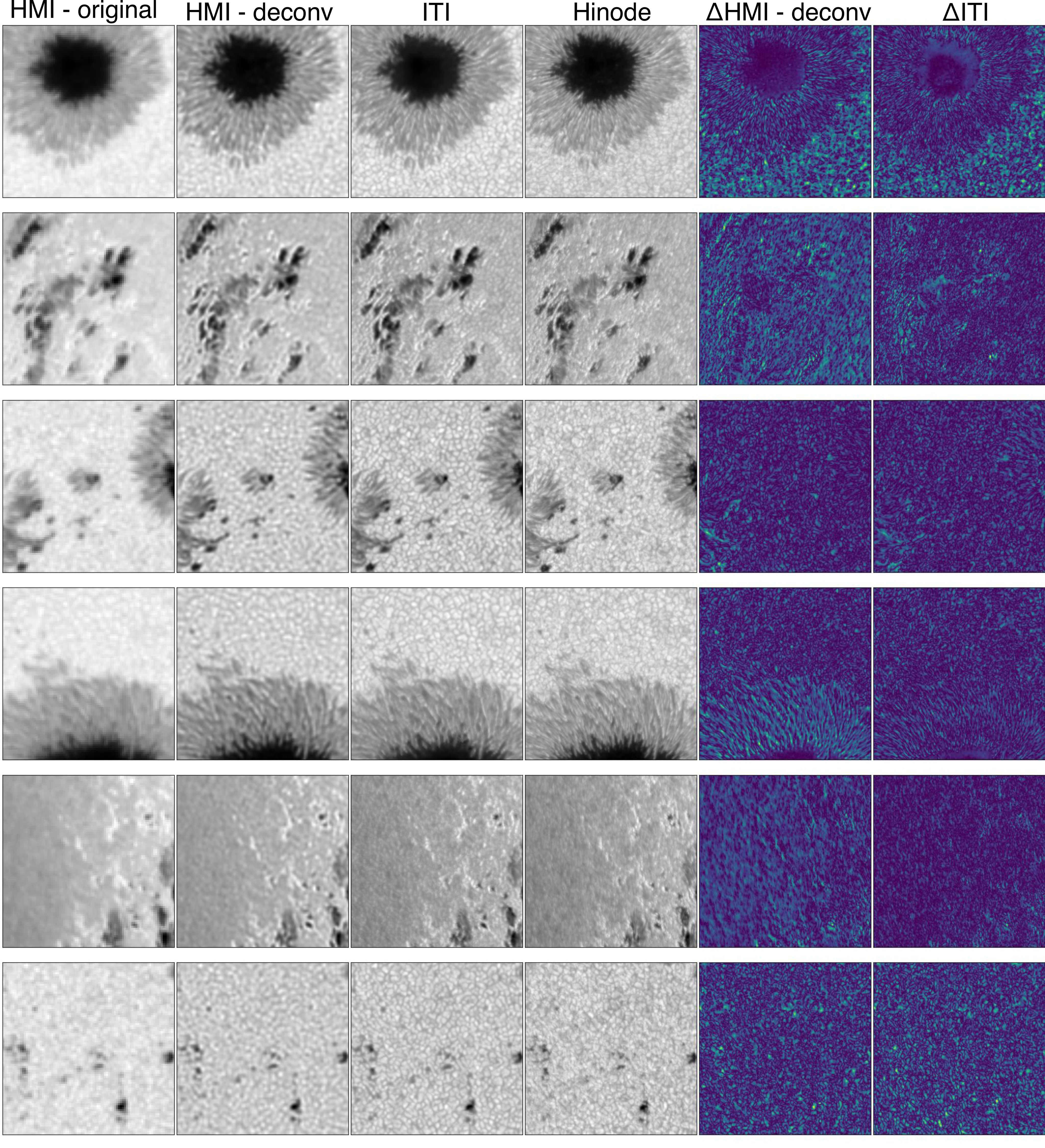}
\caption{Six examples of the HMI-to-Hinode translation. We show the original SDO/HMI observation, the deconvolved image, the ITI enhanced image, the aligned Hinode/SOT observation and the corresponding difference maps.}
\label{figure:hmi_hinode_comparison}
\end{figure}

\subsection{Variation of atmospheric degradations}

By modeling a large variety of realistic degradation effects we expect that the reconstruction model can account for multiple atmospheric degradation effects simultaneously. From the included noise term in our generator BA, multiple low-quality observations can be synthesized from a single high-quality observation. We illustrate the diversity of the generated samples by randomly selecting a high-quality image and generating six corresponding low-quality images, where we use a different noise term for each example. Four examples are shown in Supplementary Fig.~\ref{figure:kso_variation}, where panel a contains the high-quality input images and panel b the synthetic low-quality images. The samples show a large spread of different cloud distributions, both in density and position. From a comparison with Fig. \ref{figure:kso_combined} it can be seen that the synthetic clouds appear similar to real low-quality observations. The solar features are consistent with the high-quality reference (e.g., active regions, filaments).

\begin{figure}%
\centering
\includegraphics[width=\linewidth]{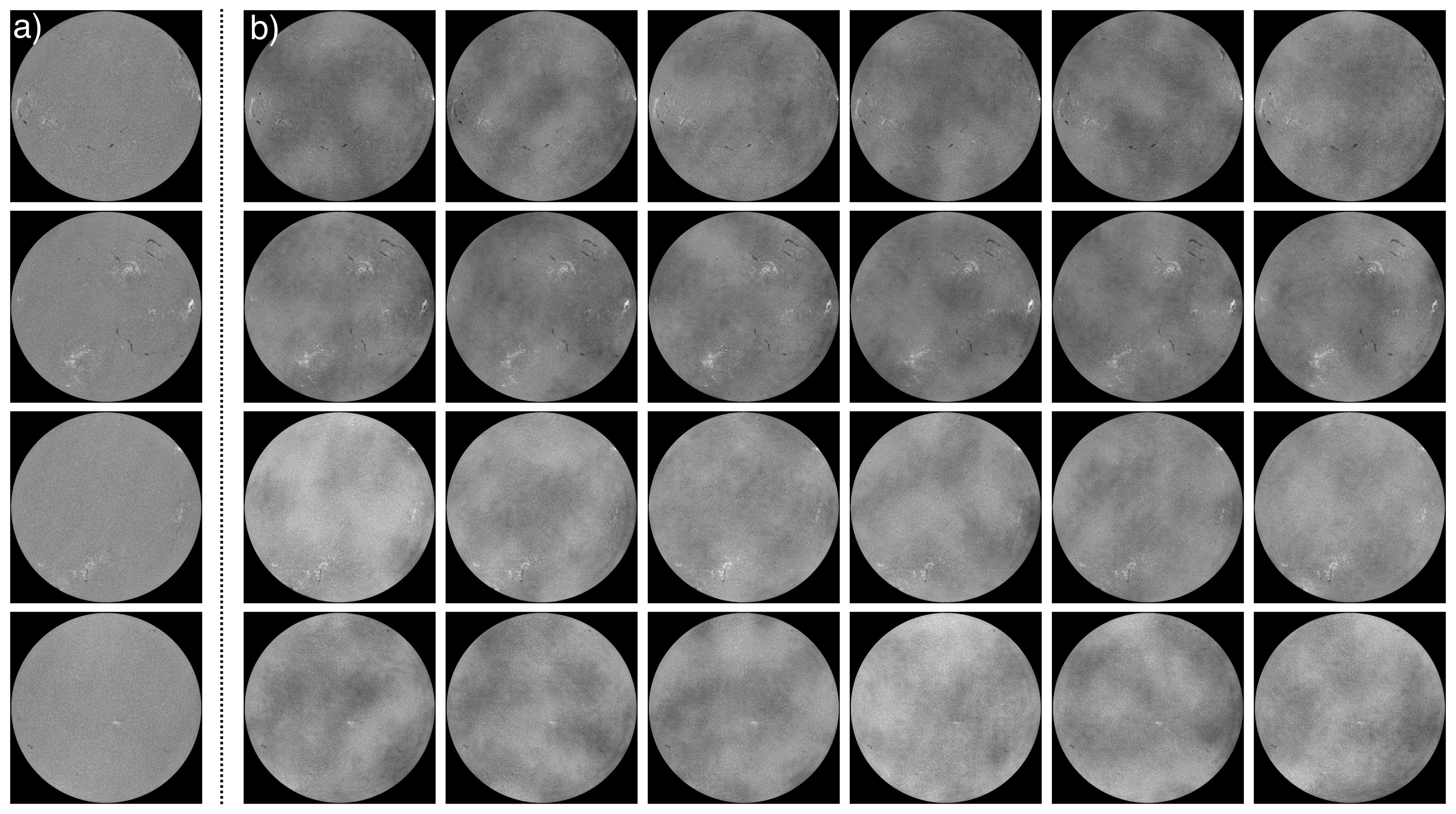}
\caption{Diversity examples of the KSO low-to-high quality translation. a) real high-quality observations. b) artificially degraded observations. The images in one row are generated from the same high-quality input image. The artificial low-quality images show a variety of realistic atmospheric effects, while the solar features are mostly unchanged.}
\label{figure:kso_variation}
\end{figure}

\subsection{Detailed comparison of enhanced photographic film observations}
\label{section:detailed_film}

A comparison between cutouts of the film scans and the ITI enhanced observations is shown in Supplementary Fig. \ref{figure:film_comparison}. All samples show a clear improvement in image sharpness, that especially affects the plage regions and leads to more distinct representation of solar filaments. As can be seen from Supplementary Fig. \ref{figure:film_comparison}b, the ITI translation also mitigates clouds, similar to the application in Sect. \ref{section:kso}.

\begin{figure}%
\centering
\includegraphics[width=0.8\linewidth]{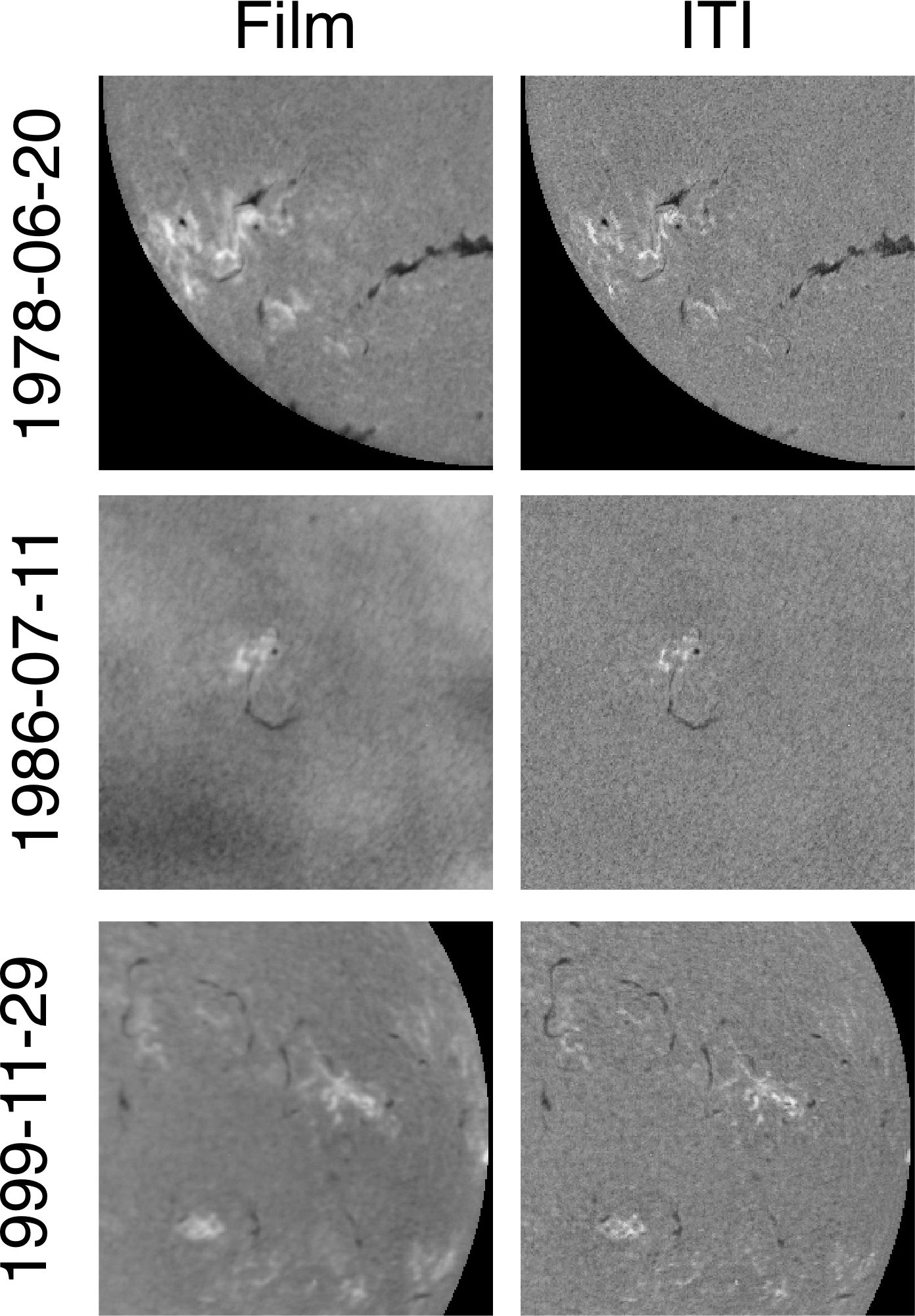}
\caption{Example of KSO film-to-CCD translation. The ITI translation results in deblurring of plage regions, sharpening of filaments and enhances the structure of quiet Sun regions. Atmospheric effects are strongly mitigated in the enhanced version (middle).}
\label{figure:film_comparison}
\end{figure}

In Supplementary Fig. \ref{figure:film_sample} we provide a comparison of a filtergram at three different scales. In Supplementary Fig. \ref{figure:film_sample}a the calibration of the image intensity and correction of large scale inhomogeneities is visible. The cutouts show a comparison between the film observation and the restored image, where we note a perceptual quality improvement.

\begin{figure}%
\centering
\includegraphics[width=0.5\linewidth]{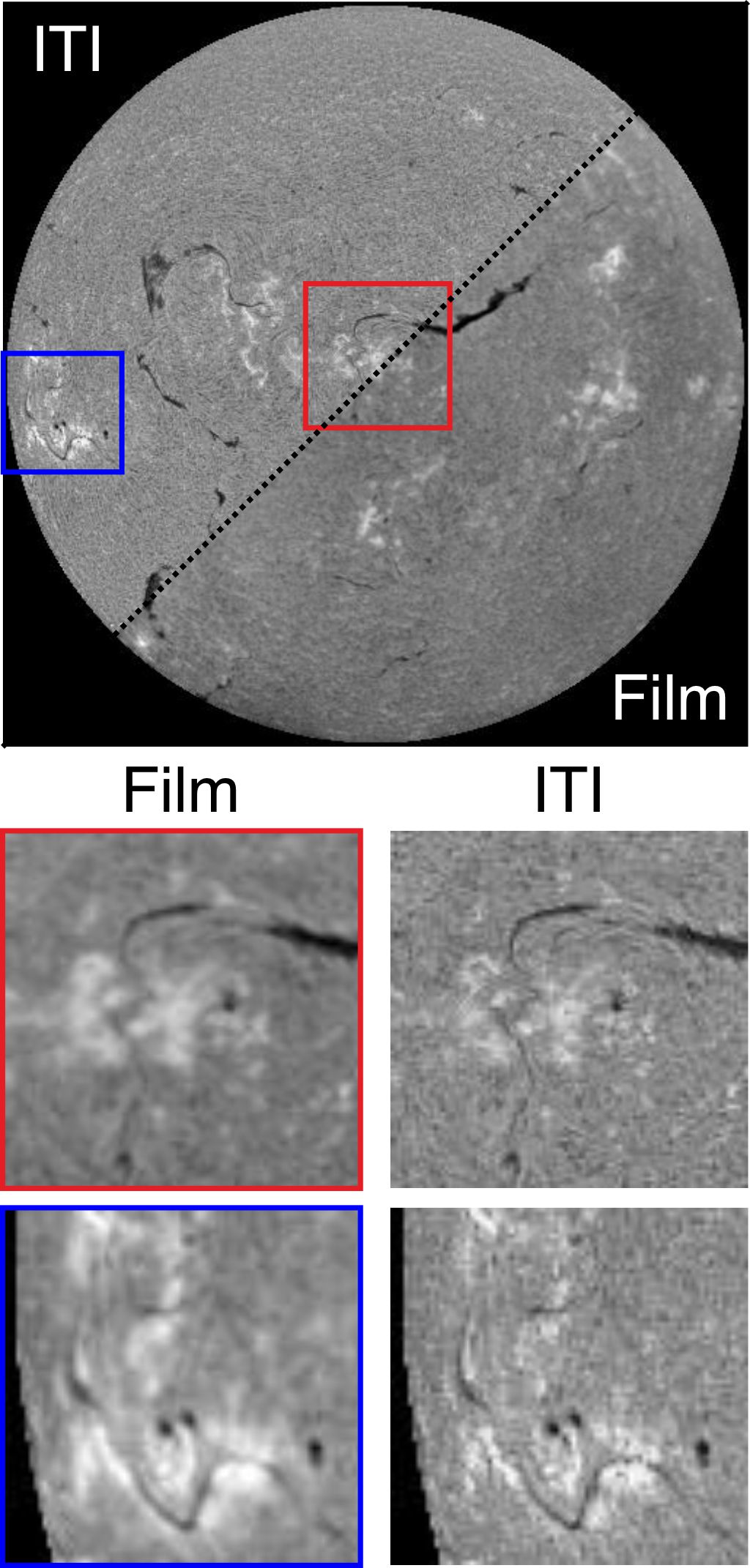}
\caption{Detailed comparison of a KSO film-to-CCD translation sample from 1992-03-12 08:51. The ITI observation shows a more homogeneous appearance at the global scale (top). The comparison of two regions shows that active regions are well reconstructed (red). At smaller scales (blue), solar features can be better identified in the ITI observations, but the limits in resolution are visible.}
\label{figure:film_sample}
\end{figure}

\subsection{Temporal stability of generated ITI magnetograms}
\label{section:consistency_stereo_mag}

We further compare the temporal stability of our method by estimating the full-disk magnetograms from 2007-01-14 to 2007-01-18 and compare the sequence to the SOHO/MDI magnetograms (Movie 3). The ITI magnetograms are overall consistent and show no large artifacts (e.g., random active regions). The active region in the ITI magnetograms appears similar to the real observation and is consistent in its position on the solar disk. The largest inconsistencies originate from variations in the estimated sunspots and we note a quality decreases close to the solar limb.


\subsection{Comparison to paired image-to-image translation}
\label{section:sdo_euv_to_mag}

Recent studies try to estimate the solar far-side magnetic field with the use of paired image translations, where first a model is trained to translate from SDO/AIA EUV filtergrams to SDO/HMI magnetograms, and apply in a second step the trained model to EUV data from STEREO \cite{kim2019solar, jeong2020ai_mag_pfss}. Here, we compare our method with the paired image translation methods. For this we consider the first step (SDO EUV $\xrightarrow{}$ magnetogram), which we can fully quantify. For the paired image-to-image translation, we use the pix2pixHD model \cite{wang2018pix2pixhd}, in line with \cite{jeong2020ai_mag_pfss, shin2020caII_to_mag,  son2021euv_to_he}. We note that our method contains the pix2pix training cycle, but it is only employed with secondary importance.

We train both methods with the same data set, using only observations from SDO, where we reduce the resolution to 512$\times$512 pixels. For pix2pix we use the standard training described in \cite{wang2018pix2pixhd}, without the VGG loss and start with 32 filters in the first convolutional layer (larger models led to divergence). The model is trained for 200 epochs. For ITI we train our model to translate from the four EUV channels (A) to the same set including the magnetogram (B). We train our model for 160.000 iterations, with the same parameter settings as in Sect. \ref{section:stereo_mag} (STEREO-To-SDO). For comparison we also train ITI to estimate the unsigned magnetic flux (no polarities). 
 
In Supplementary Table \ref{table:pix2pix_comparison} we summarize the results in terms of mean-absolute-error (MAE), MAE of the unsigned magnetic flux, and FID. The results show that the paired translation performs best for estimating the polarity of the magnetic field (MAE), while our ITI method achieves a similar performance with unpaired training. In terms of perceptual quality (FID), ITI provides the best results. Enforcing the estimation of magnetic polarities reduces the model performance for estimating the total flux distribution, as can be seen for the unsigned MAE, where the best results are obtained by the ITI model that is explicitly trained for this task.

\begin{figure}%
\centering
\includegraphics[width=\linewidth]{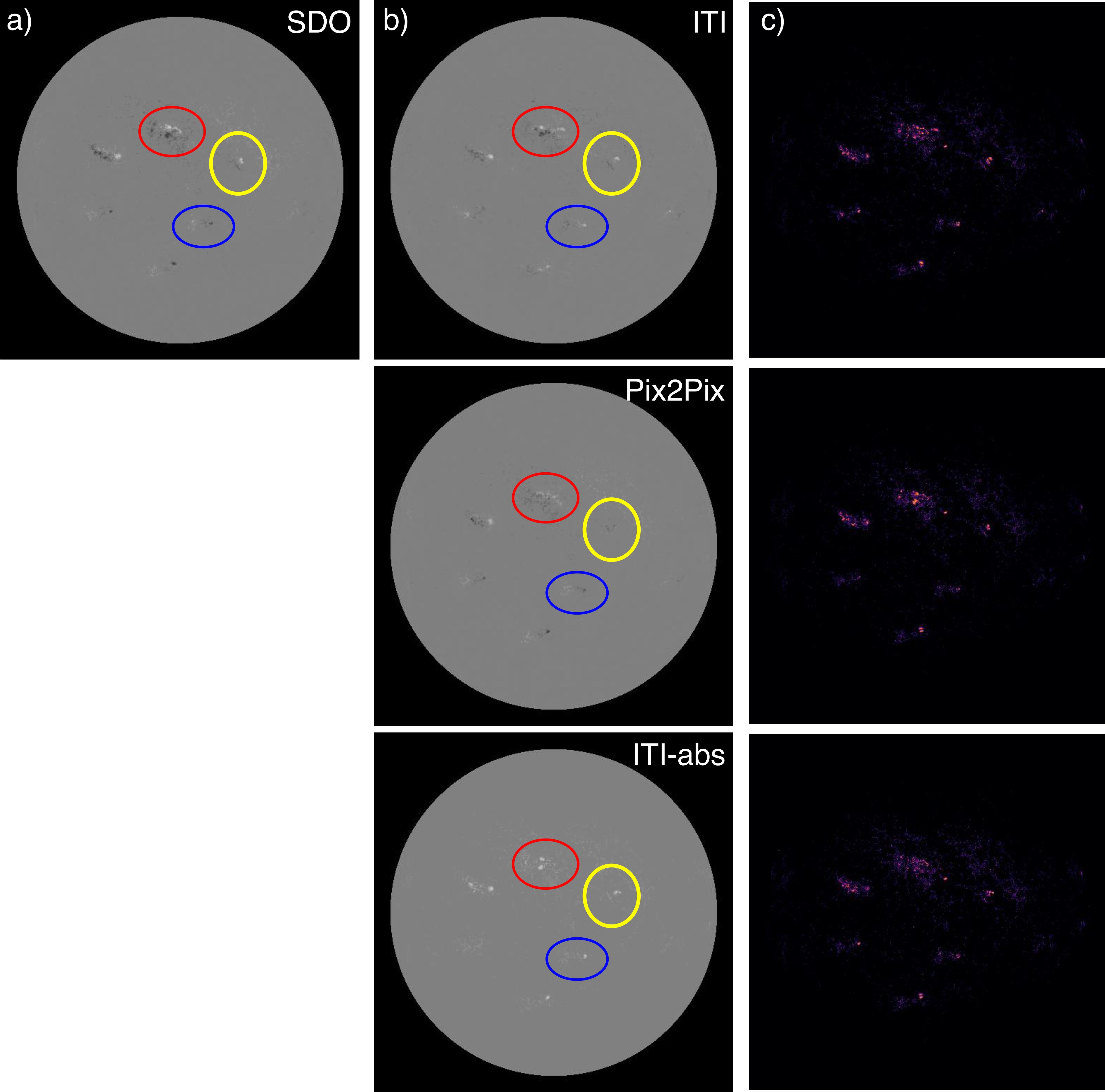}
\caption{We compare the original SDO observation with the Pix2Pix paired translations, the ITI unpaired translations, and the ITI method trained with absolute magnetic flux (2014-12-17 12:00). Column a shows the reference SDO/HMI magnetogram, column b the synthetic mangetograms, and column c the difference maps for the magnetograms of unsigned flux. All methods show larger deviations for regions of confined magnetic flux (red circle). The polarity can be confused even on global scales (blue for ITI; yellow for Pix2Pix) and also on small scales (red for both). The difference maps show that the largest error occurs due to a misalignment of estimated sunspots, as well as differences in magnetic field strength.}
\label{figure:sdo_mag}
\end{figure}

In Sect. \ref{section:stereo_soho_sdo} we used the unsigned magnetic flux for our training.  The reason for this is two fold. 1) There exists no evidence that the polarity can be directly derived from the EUV observations. 2) The polarity could be inferred from the global magnetic field configuration (i.e., Hale's Law). However, this can only be performed by using a data set of only one solar cycle. Moreover, this estimation breaks down for complex regions of strong confined flux. This is supported by Supplementary Fig. \ref{figure:sdo_mag}, where we show a comparison between Pix2Pix and the ITI results. All methods lack in estimating the correct position and strength of confined flux regions (i.e., sunspots and pores). For this comparison, none of the methods produced larger artifacts (e.g., strong magnetic flux outside of active regions).

The pix2pix approach assumes that the EUV observations are similar between SDO/AIA and STEREO/EUVI, however we found that the EUV filtergrams are vastly different (Sect. \ref{section:stereo_soho_sdo}). Since the pix2pix methods are trained for a specific image domain, it can not be assumed that comparable results are achieved when applied to data that is not within the training domain. Our method is based on domain translation, thus it explicitly learns to translate from the STEREO/EUVI to the SDO/AIA+HMI domain, from which we expect a comparable performance to this reference task.

In conclusion, our unpaired image-to-image translation approach can provide comparable results to paired translations, while being also applicable for data sets that have no spatial or temporal alignment.

\begin{table}
\caption{Evaluation of the SDO EUV to magnetogram training task for the Pix2Pix method and ITI. For all metrics lower values correspond to a better performance.}             
\label{table:pix2pix_comparison}      
\centering                          
\begin{tabular}{| l || c c c |}        
\hline
Method & MAE [G] $\downarrow$ & MAE - unsigned [G] $\downarrow$ & FID $\downarrow$ \\    
\hline                        
   Pix2Pix & \textbf{8.5} & 6.6 & 4.1 \\
   ITI & 9.6 & 7.3 & \textbf{1.6} \\
   ITI-abs & - & \textbf{6.1} & - \\
\hline                                   
\end{tabular}
\end{table}

\end{appendix}
\end{document}